\documentclass[12pt]{article}
\usepackage{amsmath,amssymb,amsfonts,fancyhdr,fancybox,graphics,graphicx,psfrag,calc,color,hyperref,times,color,verbatim,lscape,empheq}
\usepackage{MnSymbol}
\usepackage{enumerate}
\usepackage{relsize}
\usepackage{url} 
\usepackage{multirow}
\usepackage{multicol}
\usepackage{rotating}
\usepackage{relsize}
\usepackage{bm}
\usepackage{hyperref}
\usepackage{multicol}
\usepackage{setspace}
\usepackage{enumerate}
\usepackage[authoryear,sort]{natbib}
\usepackage{footnote}
\usepackage{authblk}
\usepackage{mymacros}

\def\boxit#1{\vbox{\hrule\hbox{\vrule\kern6pt\vbox{\kern6pt#1\kern6pt}\kern6pt\vrule}\hrule}}

\graphicspath{{figure/}}

\begin{document}

\title{\bf Estimation and Selection Properties of the LAD Fused Lasso
Signal Approximator}
\date{}

\author{\begin{tabular}{c} Xiaoli Gao\footnote{Correspondence: 106 Petty Building, Greensboro, NC 27412. Email: x\_gao2@uncg.edu}\\
\emph{Department of Mathematics and Statistics}\\
\emph{University of North Carolina at Greensboro}
\end{tabular}}
\titlepage

\maketitle

\begin{center}
\begin{minipage}{130mm}
\begin{center}{\bf Abstract}\end{center}
%

  The fused lasso is an important method for signal processing
  when the hidden signals are sparse and blocky. It is often
  used in combination with the squared loss function.
  However, the squared loss is not suitable for heavy tail error distributions nor is robust against outliers which arise
  often in practice.  The
  least absolute deviations (LAD) loss provides a robust alternative
  to the squared loss.
 In this paper, we study the asymptotic
 properties of the  fused lasso estimator with the LAD loss for
 signal approximation. We refer to this estimator as the LAD fused
 lasso signal approximator, or LAD-FLSA.
 We investigate the  estimation consistency properties of the LAD-FLSA
 and provide sufficient conditions under which the LAD-FLSA  is
 sign consistent. We also construct an unbiased estimator for the  degrees of freedom
 of the LAD-FLSA  for any given tuning parameters.
 Both simulation studies and real data analysis are conducted to illustrate the performance of the LAD-FLSA and the effect of the unbiased estimator of the degrees of freedom.
%
%
%
%
%
%
%
%
%
%

  \end{minipage}
\end{center}


\newpage
\setcounter{equation}{0}

\newpage
\setcounter{equation}{0}


\section{Introduction}



High-dimensional data arise in many fields including signal
processing, image de-noising
and genomic and genetic studies. When a model is sparse and has certain known structures, penalized methods have been
widely used  since they can incorporate
known structures into penalty functions in a natural way and can do estimation and variable selection simultaneously.
A biological example is the analysis of copy-number
variations  in a human genome.  In this problem,  we are interested in
detecting the changes in copy numbers based on
data from comparative genomic hybridization (CGH) arrays.
For instance, Snijders et al. (2001) studied a CGH array consisting of $2400$ bacterial artificial chromosome (BAC) clones, where
the log base 2 intensity ratios at all clones are measured. In Figure 1, we plot a sample CGH copy number data on chromosomes 1--4 from cell line GM 13330. The data set has two characteristics:
1) there are only a small number of chromosomal locations where
the copy numbers of genes change, that is, the underlying signals are
sparse;
2)  the adjacent markers tend to have similar observations, i.e., the signals are blocky.


In a signal approximation model with sample size $n$, the $i$th observation $y_i$ is considered
to be
a realization of the true signal  $\mu^{0}_{i}$ plus random noise $\veps_i$,
 \bel{signal approximation model}
  y_i=\mu^{0}_{i} +\veps_i, \ \ i=1, \cdots, n.
\eel
In many cases, the true signal vector $\bmu^0=(\mu^0_{1},\cdots,\mu^0_{n})'$ is
both blocky and sparse, meaning that the intensities of true signals within each block are  the same and most blocks consist of zero signals.
The goal here is to find a solution  not only to recover all
the changes points, but also to identify the nonzero blocks.
We can use the lasso penalty to enforce a sparse solution by
penalizing the $\ell_1$ norm of the signals
$\|\bmu\|_1\equiv\sum_{i=1}^n |\mu_i|$,
and use the fusion (total variation) penalty to enforce
a blocky solution by penalizing
  $\|\bmu\|_{\rm TV}\equiv\sum_{i=2}^n |\mu_i-\mu_{i-1}|$.
The combination of these two penalties results in the fused lasso (FL) penalty (Tibshirani et al., 2005).

For detecting changing points in copy number variations, Tibshirani and Wang (2008) proposed to use the fused lasso with a squares  loss function. We refer to this approach as the least squares fused lasso signal
approximator (LS-FLSA). For  $\lamn = (\lmone, \lmtwo)$, the LS-FLSA  seeks to find
 $\hbmu^{\ell_2}_n(\lamn)=
 (\hmu^{\ell_2}_1(\lamn),\cdots,\hmu^{\ell_2}_n(\lamn))'$ that minimizes
\bel{tw-ls flsa model}
\sum_{i=1}^n(y_i-\mu_i)^2
+\lmone\sum_{i=1}^n
|\mu_i|+\lmtwo\sum_{i=2}^n|\mu_i-\mu_{i-1}|,  \eel
 where $\lmone$ and $\lmtwo$ are two nonnegative penalty parameters.

Recently,
Rinaldo (2009) studied the selection properties
of the LS-FLSA and adaptive LS-FLSA under the block partition assumption
in the underlying signal.
Several authors have also studied the properties of related procedures. For example, Mammen and van de Geer (1997) studied the rate of convergence in bounded variation function classes of the parameter functions; Harchaoui and L\'{e}vy-Leduc (2010) investigated the estimation properties of change points using
the total variation penalty;
Boysen et al. (2009) studied the asymptotic properties for jump-penalized least-squares regression aiming at approximating a regression function by piecewise-constant functions.
These studies significantly advanced our understanding of LS-FLSA
or fusion penalized LS methods in the context signal detection or
nonparametric estimation.
However, all these results are obtained for methods with the least squares loss and/or require normality assumption on the errors.
The LS-FLSA is easily affected by outliers which arise often
in practice, for example, in CGH copy number variation data.



A more robust  fused lasso signal approximator
can be constructed by using the LAD loss, which we shall refer to as LAD-FLSA. For any given $\lamn=(\lmone, \lmtwo)$, the LAD-FLSA is defined as
 \bel{lad-flsa model}
\hbmu_n(\lamn)=\argmin_{\bmu\in \cR_n}\left\{\sum_{i=1}^n|y_i-\mu_i|
+\lmone\sum_{i=1}^n
|\mu_i|+\lmtwo\sum_{i=2}^n|\mu_i-\mu_{i-1}|\right\}.  \eel
The convex minimizer $\hbmu_n$ in  \eqref{lad-flsa model}
has  been applied to detect
copy number variation breakpoints in Gao and Huang (2010a). However,
its  theoretic properties of  have not  been studied.

In this paper, we seek to answer the following questions about
the LAD-FLSA: (1) how  close  $\hbmu_n^{\ell_1}$
can be to the true model $\bmu^0$ asymptotically?
(2) how accurately  $\hbmu_n^{\ell_1}$ can
recover the true nonzero blocks with a large probability when $\bmu^0$ is both sparse and blocky? (3) what is the degrees of freedom of
LAD-FLSA?
The contributions of this paper are as follows.

\begin{itemize}
\item
We show that the LAD-FLSA
is rate consistent if the number of blocks is relatively small.

\item
We provide sufficient conditions under which the LAD-FLSA
is able to recover the block patterns and
distinguish nonzero blocks from zero ones correctly with high probability. That is, the LAD-FLSA can determine all the jumps, identify all the nonzero blocks, and also distinguish the positive signals
from negative ones under some conditions.

\item In terms of model complexity measures, we justify that the number of nonzero blocks generated from
a LAD-FLSA estimate is an unbiased estimator of the degrees of freedom of the LAD-FLSA.

\item Without the assumption of Gaussian or sub-Gaussian random error, our studies can be widely applied for signal detection in signal processing when  random noises  do not follow nice
distributions or the signal-to-noise ratios are large.
\end{itemize}

The rest of the paper is organized as follows.
We list some notations in Section 2.
We study the estimation consistency and sign consistency properties of
the LAD-FLSA  respectively in Section 3 and 4.
 In Section 5 we derive an unbiased estimator
 of the degrees of freedom  of the LAD-FLSA.
In Section 6 we conduct simulation studies and real data analysis
to demonstrate the
performance of the LAD-FLSA.
We also verify the effect of unbiased estimator of the degrees of freedom numerically in this section.
Section 7 concludes the paper with some
discussions.
All the technical proofs are postponed to the Appendix.

 \section{Preliminaries}
 Suppose the true model $\bmu^0=(\mu^0_1,\cdots,\mu^0_n)'$ includes $J_0$ blocks
 and there exists a unique vector $\bnu^0=(\nu^0_1, \cdots,\nu^0_{J_0})$ such that
 \bel{truemodel}
\mu_i^0=\sum_{j=1}^{J_0} \nu_j^0~ I(i \in \cB_j^0),
\eel
where  $\{\cB_1^0, \cdots, \cB_{J_0}^0\}$ is a mutually exclusive  block partition of $\{1,\cdots, n\}$ generated from $\bmu^0$.
Based on the block partition, we can rewrite $\{1,\cdots,n\}$ as
$\{i_0,\cdots,i_1-1, i_1,\cdots, i_2-1, \cdots,i_{J_0-1},\cdots, i_{J_0}-1\}$, where
$1=i_0<i_1<\cdots<i_{J_0-1} \le i_{J_0}-1=n$ and $\{i_{j-1},\cdots,i_j-1\}$ gives
the $j$th block set $\cB_j^0$.
We denote the jump set in the true model as $\cJ^0$.
Then $\cJ^0=\{i_1,\cdots,i_{J_0-1}\}$ and
$J_0=|\cJ^0|+1$, where $|\cJ^0|$ is the
 the cardinality of $\cJ^0$.
Let $\cK^0=\cK(\bmu^0)=\{1\le j\le J_0: \nu_j^0\neq 0\}$
be the set of nonzero blocks of $\bmu^0$ and the number of nonzero blocks
$K_0=|\cK^0|$.
We now list the following notations that will be used throughout the
 paper, some of which are adopted from Rinaldo (2009).
\begin{itemize}
\item For the true model $\bmu^0$ defined  in \eqref{truemodel}, we introduce the following notations (I--IV):
 \begin{itemize}
 \item [(I)]  $b_j^0=|\cB_j^0|$, the size of the block set $\cB_j^0$ for $1\le j\le J_0$;
  \item [(II)]  $b^0_{\min}=\min_{1\le j\le J_0} b_j^0$, the smallest block size;
  \item [(III)]  $a_n=\min_{i\in \cJ^0} |\mu_i^0-\mu_{i-1}^0|$, the smallest jump;
  \item [(IV)]  $\rho_n=\min_{j\in\cK^0} |\nu_j^0|$, the smallest nonzero signal intensity.
  \end{itemize}
\item  Corresponding notations are analogous to  a LAD-FLSA estimate $\hbmu_n$  in
    \eqref{lad-flsa model} as follows:
  \begin{itemize}
    \item [(V)]  $\widehat \cJ=\cJ(\hbmu_n)$, $\Jhat=J(\hbmu_n)$, $ \hcB_j=\cB_j(\hbmu_n)$
      with $\widehat b_j=|\hcB_j|$  and
  $\hnu_j=\nu_j(\hbmu_n)$ for $1\le j\le \hcBJhat$ are the estimated jump set, number of blocks,
   block partitions of $\{1,\cdots, n\}$  with corresponding block size  and the associated unique
  vector generated
    from  $\hbmu_n$;
      \item [(VI)]  $\widehat{\cK}=\cK(\hbmu_n)=\{1\le j\le \Jhat: \hnu_j\neq 0\}$ is the set of estimated nonzero blocks
   and $\Khat=|\widehat{\cK}|$.
     \end{itemize}
 \end{itemize}

 \section{Estimation consistency of LAD-FLSA estimators}\label{est-con}
In this section, we study the estimation consistency of the LAD-FLSA  $\hbmu_n$.
We first consider the following conditions:
 \begin{itemize}
 \item[(A1)] Error assumption: random errors $\veps_i$'s in model
 \eqref{signal approximation model} are independent and
identically distributed with median $0$, and have a density $f$ that is continuous and positive 
in a neighborhood of $0$.
\item[(A2)] Block number assumption: the true block number $J_0<M_1\Lambda_n$ for a constant $M_1 > 0$, where
$\Lambda_n=\max\{16n/(\lmtwo^2-2n^2\lmone^2), n/(\lmtwo-n\lmone)\}+1$ with $\lmtwo^2>2n\lmone^2$.
\end{itemize}
In (A1), we only require that the random errors have a density that
is continuous and positive in neighborhood of zero and have median
zero. This is a much weaker condition than the  Gaussian random error
assumption required in Harchaoui and L\'{e}vy-Leduc (2010) and Rinaldo (2009). Indeed, (A1) allows all heavy-tail distributions of the errors,
including the Cauchy distribution whose moments do not exist.
Condition (A2) requires that the number of blocks in the underlying model can increase
with $n$, but at a slower rate than $O(\Lambda_n)$.
As a by-product, the tuning parameter for jumps, $\lmtwo$, grows much faster that the tuning parameter
for signals intensities, $\lmone$.  It is a reasonable  assumption since the true model is block-wise, that
 is, the number of nonzero jumps can be much smaller than the number of the nonzero signals.
For example, if the number of jumps is  $O((\log(n))^{1/2})$,
then we can let $\lmtwo=n^{1/2}(\log(n))^{-1/4}$ and $\lmone=n^{-1}$.


In order to study the asymptotic properties  of the LAD-FLSA estimator $\hbmu_n$ in
  \eqref{lad-flsa model}, we first  investigate its LS-FLSA  approximation,
  \bel{ls-flsa model}
 \tbmu_n(\lamn)=\argmin_{\bmu}\left\{\sum_{i=1}^n(z_i-(f(0))^{1/2}\mu_i)^2
 +\lmone\sum_{i=1}^n |\mu_i| +\lmtwo \sum_{i=2}^n|\mu_i-\mu_{i-1}|\right\},
 \eel
 where $z_i=(f(0))^{1/2}\mu^0_i+\eta_i$ with $\eta_i=(4f(0))^{-1/2}\sgn(\veps_i)$ for $1\le i\le n$
 consist of  pseudo-signal data.
 Thus, all estimates listed in (V--VI)  can be
analogous to the corresponding ones
generated from $\tbmu_n$.
    We now provide  some rate upper bounds for the number of blocks
    generated from $\tbmu_n$ in \eqref{ls-flsa model} and $\hbmu_n$ in \eqref{lad-flsa model}, respectively.

\begin{lemma}\label{lemma dmax} Under (A1), we have
 (i) $\widetilde J\le 16n/(\lmtwo^2-2n^2\lmone^2)+1$,
provided $\lmtwo^2>2n^2\lmone^2$ and
(ii) $\widehat J\le n/(\lmtwo-n\lmone)+1$, provided
 $\lmtwo>n\lmone$.
 In addition, suppose (A2) holds, we also have
 (iii) $\widetilde J+\widehat J+J_0<(M_1+2)\Lambda_n$,
 where both $M_1$ and $\Lambda_n$ are defined in (A2).
\end{lemma}
The proof of Lemma \ref{lemma dmax} is given in the Appendix.
   Lemma \ref{lemma dmax} gives upper bounds for the number of blocks associated with  $\tbmu_n$ and $\hbmu_n$.
We can interpret bounds in (i) and (ii) as the maximal
dimension of any linear space  where
  $\hbmu_n$ and $\tbmu_n$ may belong, respectively.
  Similarly, (iii) provides us an unified rate upper bound
  for the dimension of any linear space to which
 $\hbmu_n$, $\tbmu_n$ and $\bmu^0$ can belong.
  Lemma \ref{lemma dmax}  is useful in obtaining   the estimation consistencies of
   $\hbmu_n$ and $\tbmu_n$. Furthermore,
    it is important to notice that those upper bounds in Lemma \ref{lemma dmax}
 are mainly affected
by the rate of  $\lmtwo$, which  is reasonable since the number of jumps  in an FLSA model is mainly determined
by $\lmtwo$.

Denote   $\|\bmu\|_n^2=\sum_{i=1}^n \mu_i^2 /n$
and $\|\bmu\|_2^2=\sum_{i=1}^n \mu_i^2.$ Below we present the  estimation properties of
$\tbmu_n$ in \eqref{ls-flsa model}.

 \begin{lemma} \label{ls-flsa consistency}
Suppose (A1-A2) hold.   
Then there exists a constant $0<c<1$,   such that
 $$\bP\left(\|\tbmu_n-\bmu^0\|_n \ge \alpha_n \right) \le  \Lambda_n \exp\{ \Lambda_n\log n -(1-c)^2(f(0)/2) n\alpha_n^2\},$$
 where
 $\Lambda_n$ is defined in (A2) and
 $\alpha_n=1/(c\sqrt{f(0)}) [\lmone+2\lmtwo
     +((M_1+1)\Lambda_n/n)^{1/2}]$.
 Furthermore, if we let $\alpha_n=(2M_2 \Lambda_n (\log n)/n)^{1/2}$ and
choose $\lmone$ and $\lmtwo$ such that
$\lmone+2\lmtwo=c\sqrt{f(0)}\alpha_n-((M_1+1)\Lambda_n/n)^{1/2}$
 for a constant $M_2>1/(f(0)(1-c)^{2}$, then
 $$ \bP\left(\|\tbmu_n-\bmu^0\|_n \ge \alpha_n\right)
\le \Lambda_n n^{\{1-M_2f(0)(1-c)^2\}\Lambda_n}.$$
  \end{lemma}
The proof of Lemma 2 is given in the Appendix.
 Lemma \ref{ls-flsa consistency} gives  us the estimation consistency result for
a pseudo LS-FLSA  $\tbmu_n$ (using  pseudo data $z_i$'s
and bounded noises $\eta_i$'s).
It is worthwhile to point out that even though we only report the consistency result for
 a pseudo LS-FLSA estimator $\tbmu_n$ with 
  bounded noises $\eta_i$'s  in Lemma 2,
we can obtain a similar consistency result for the regular LS-FLSA
 estimator \eqref{tw-ls flsa model} under the assumption of Gaussian noises without much extra work.
Thus, the estimation consistency properties
 of the LS signal approximator with  the total variation penalty
in Harchaoui and L\'{e}vy-Leduc (2010)
can also be obtained from Lemma 2 by taking $\lmtwo=0$  and $\Lambda_n=K_{\max}$.

The consistency result of $\tbmu_n$ in Lemma  \ref{ls-flsa consistency} plays an important role in deriving the corresponding
estimation consistency result  of $\hbmu_n$  in the following Theorem \ref{lad-flsa consistency}.
\begin{theorem}\label{lad-flsa consistency}
Suppose (A1) and (A2) hold. Then
there exists a constant $0<c<1$   such that
 $$\bP\left(\|\hbmu_n-\bmu^0\|_n \ge \gamma_n \right)
 \le  \Lambda_n \exp\{ \Lambda_n\log n -(1-c)^2(f(0)/8) n\gamma_n^2\}
                     +(8/f(0))(\Lambda_n/(n\gamma_n^2))^{1/2},$$
 where
 $\Lambda_n$ is defined in (A2) and
 $\gamma_n=2/(c\sqrt{f(0)}) [\lmone+2\lmtwo +((M_1+1)\Lambda_n/n)^{1/2}]$.

Furthermore, if we let $\gamma_n=(8M_3 \Lambda_n (\log n)/n)^{1/2}$
for a constant $M_3>1/(f(0)(1-c)^2)$
and choose $\lmone$ and $\lmtwo$ such that
$\lmone+2\lmtwo=(c\sqrt{f(0)}/2)\gamma_n-((M_1+1)\Lambda_n/n)^{1/2}$
, then
$$ \bP\left(\|\hbmu_n-\bmu^0\|_n \ge \gamma_n\right)
\le \Lambda_n n^{-\{M_3f(0)(1-c)^2-1\} \Lambda_n}+O\left(1/\sqrt{\log n}\right).$$
 \end{theorem}
The proof of Theorem \ref{lad-flsa consistency}
is given in the Appendix.
Theorem \ref{lad-flsa consistency} implies that
 the LAD-FLSA  $\hbmu_n$ can be consistent for estimating $\bmu^0$ at
the rate of $O\left(\Lambda_n (\log n)/ n)^{1/2}\right)$.
Furthermore, if the number of blocks in true signals is bounded,
 the  rate of convergence can be stated more explicitly
as in the following
 Corollary~\ref{corollary con}.
   \begin{corollary}\label{corollary con}
   Suppose (A1) holds and there exists $J_{\max}>0$ such that $J_0 \le J_{\max}$.
  Then
   $$ \bP\left( \{\max(\Jhat,\widetilde J)<J_{\max}\}
                \cap
                 \{\|\hbmu_n-\bmu^0\|_n \ge \theta_n\} \right)
\le J_{\max} n^{-c_{2M} J_{\max}}+O\left(1/\sqrt{\log n}\right)$$
for $\theta_n=(8M J_{\max} (\log n)/n)^{1/2}$ and
  $\lmone+2\lmtwo=(c_{1M} J_{\max} (\log n)/n)^{1/2}-(J_{\max}/n)^{1/2}$.
Here  $M>1/((1-c)^2(f(0))$ is a constant,   $c_{1M}=(2Mc^2(f(0))^{1/2}$
  and $c_{2M}=f(0)M(1-c)^2-1$.
      \end{corollary}

Corollary \ref{corollary con} says that the $\hbmu_n$ is consistent
 for estimating $\bmu^0$ at the rate $O\left( ( J_{\max} (\log n)/n)^{1/2} \right)$ if the numbers  of both true and
estimated jumps are bounded above. This convergence rate can be compared to
 $n^{-1/2}$, which is argued by
 Yao and Yu (1989) to be optimal for LS estimators of the levels of a step function.
 Notice that if $\lim_{n\to\infty}\bP(\widehat\cJ=\cJ^0)=1$, then
 $\sum_{i=1}^n(\hmu_i-\mu^0_i)^2=\sum_{j=1}^{J_0}b_j^0(\hnu_j-\nu^0_j)^2
  \ge b_{\min}^0\sum_{j=1}^{J_0}(\hnu_j-\nu^0_j)^2$ for large  $n$ almost surely.
 Thus Corollary \ref{corollary con} implies that,  for large $n$
 \bel{v-con}
   \bP\left(\{\widehat\cJ=\cJ^0\} \cap \{\|\hbnu_n-\bnu^0\|_2 \ge (8M J_{\max} (\log n)/ b_{\min}^0)^{1/2}\}\right)
\le J_{\max} n^{-\{f(0)M(1-c)^2-1\}J_{\max}},\eel where the convergence rate is affected by $b^0_{\min}$.
Therefore,  $\hbnu_n$ can converge to $\bnu^0$ in $\ell_2$ norm at rate
$O\left((J_{\max}(\log n)/b_{\min}^0)^{1/2}\right)$.
In other words, a block estimator $\hbnu_n$ can converge faster to the true model $\bnu_0$ with larger block size.

 \section{Block sign consistency of LAD-FLSA}\label{sign con}

In this section,
we study the sign consistency of the LAD-FLSA.
The sign consistency has been studied  by Zhao and Yu (2006)
and Gao and Huang (2010b) for both the LS-Lasso and LAD-Lasso in high-dimensional linear regression settings. It is a stronger result than variable selection consistency since it not only requires that  variables to be selected correctly, but also their signs are estimated correctly with high probability.

In light of  the block structure in the hidden signals,
  we consider  the selection consistency and sign consistency
 for jumps and blocks separately.
 \begin{definition}\label{jump selection consistency}
 $\hbmu_n$ is {\it jump selection consistent} if
$$
\lim_{n\to \infty} \bP\left(\{\Jhat = J_0\} \bigcap\{ \cap_{1\le j\le J_0}\{\hcB_j=\cB_j^0 \}\}\right)=1.
$$
\end{definition}
\begin{definition}\label{jump sign consistency}
 $\hbmu_n$ is {\it jump sign consistent} if
$$
\lim_{n\to \infty} \bP\left(\{\widehat \cJ = \cJ^0\}
      \bigcap \{\sgn(\hmu_i-\hmu_{i-1} )=\sgn(\mu_i^0-\mu_{i-1}^0) , \forall i\in \cJ^0\}\right)=1.
$$
\end{definition}

Definition \ref{jump selection consistency} requires that $\hbmu_n$ can  partition the signals into
 blocks correctly with probability converging to one.
Definition \ref{jump sign consistency} is a stronger requirement since
it requires that
$\hbmu_n$ finds not only all the jumps,
but also the jump directions (up/down)  correctly.
A jump selection consistent estimator can
recover the jumps set $\cJ^0$ correctly with high probability,
but does not tell us which blocks have
nonzero intensities. In other words,
 there may  exist $\delta>0$ and  $1\le j\le J_0$ such that
$P(\{\hnu_j\neq 0\} \cap \{\nu_j^0=0\})>\delta$ for a jump sign consistent $\hbmu_n$.

We now define the block selection consistency
and  the block sign consistency
in Definitions \ref{block selection consistency} and \ref{block sign consistency}.
The latter is a stronger definition since it requires the
signs of the signals to be recovered correctly.

\begin{definition}\label{block selection consistency}
 $\hbmu_n$ is {\it block selection consistent} if
$$
\lim_{n\to \infty} \bP\left( \{\widehat \cJ = \cJ^0 \}
\bigcap \{\widehat{\cK} = \cK^0\}\right)=1.
$$
\end{definition}
\begin{definition}\label{block sign consistency}
 $\hbmu_n$ is {\it block sign consistent} if
$$
\lim_{n\to \infty} \bP\left(\{\widehat \cJ = \cJ^0\}
\bigcap \{\widehat{\cK} = \cK^0\}
 \bigcap \{\sgn(\hnu_j)=\sgn(\nu_j^0), \forall j\in J_0\}\right)=1.
$$
\end{definition}

\subsection{Jump selection consistency}
For $\lmone=0$, a LAD-FLSA
becomes a LAD signal approximator  using only the total
variation penalty (LAD-FSA), defined as
  \bel{LAD-FSA}
  \hbmu_n^{\rm F}(\lm_{2n})= \hbmu_n^{\rm FL}(0,\lm_{2n})=\argmin\left\{\sum_{i=1}^n|y_i-\mu_i|
+\lambda_{2n}\sum_{i=2}^n|\mu_i-\mu_{i-1}|\right\}.
  \eel
Suppose $\hbmu_n^{\rm F}(\lm_{2n})$ can do the block partition correctly.
Then we  expect to sort out those nonzero blocks  by increasing $\lmone$ slowly from $0$.
So we first investigate the jump selection consistency
 of $\hbmu_n^{\rm F}(\lm_{2n})$.
Below we list some conditions on the smallest value of
true jumps and smallest size of the true blocks in  model \eqref{signal approximation model}
and \eqref{truemodel} for the jump sign consistency.
Recall that $b_{\min}^0$ and $a_n$ are defined in II and III in Section 2.
 \begin{itemize}
\item[(B1)]  (a) $\lmtwo\to \infty$;
     (b) there exists a $\delta>0$, such that $ \lmtwo (\log(n-J_0))^{-1/2}>(1+\delta)/2$.
\item [(B2)]  (a) 
$ (b^0_{\min})^{1/2}a_n\to \infty$;
               (b) there exists $\delta>0$, such that
$ (b^0_{\min}/\log(J_0))^{1/2}a_n>3(1+\delta)/(\sqrt{2}f(0))$ for sufficiently large $n$.
\item [(B3)]  $\lmtwo<(f(0)/3) b_{\min}^0 a_n$ for sufficiently large $n$.
\end{itemize}
Here (B1) and (B3) indicate that $\lmtwo$  increases with $n$
 faster than $O\left(\log(n-J_0)^{1/2}\right)$ but
  slower than $O(b_{\min}^0 a_n)$.
(B2-a)  requires that either the smallest jump or the  smallest size of all blocks
in the true model should be
large enough so that $\{1,\cdots,n\}$ can be partitioned into different blocks  correctly.
(B2-b) strengthes (B1-a) by providing a lower bound.
Conditions (B1-B3) provide us some helpful information in finding
an optimal tuning parameter
 in model  \eqref{LAD-FSA}.
     When the above conditions are satisfied, the LAD-FSA estimator $\hbmu_n^{\rm F}(\lm_{2n})$ can
group  all signals into different blocks correctly with a large probability.
 \begin{theorem}\label{theorem-consistency-fsa}
 Consider the signal approximation model \eqref{signal approximation model}
 with the true model \eqref{truemodel}.
  A LAD-FSA estimator $\hbmu_n^F(\lm_{2n})$ is jump sign consistent under (A1) and (B1-B3).
\end{theorem}
The proof of Theorem \ref{theorem-consistency-fsa} is postponed to the Appendix.
 Theorem~\ref{theorem-consistency-fsa} tells us that we can apply a LAD-FSA approach to  recover not only the true jumps, but also their signs
correctly  with high probability if the true hidden signal vector is blocky
and the tuning parameter $\lmtwo$ is chosen appropriately.

\subsection{Block selection consistency}

We have seen that a LAD-FSA solution can
 be  jump selection consistent
 to the  blocky hidden signal vector
  under some conditions.
 In many cases, the true signal vector
 includes some zero blocks, which cannot be
 separated from nonzero ones using the LAD-FSA
approach since
 the total variation penalty only shrinks adjacent differences but not signals
  themselves.
The additional lasso
 penalty of FLSA can
force the estimates of some block intensities to be exactly zero.
 We are interested in
 finding a LAD-FLSA solution to  not only recover the true jumps, but also find
  the zero blocks and keep only the nonzero ones
   with a large probability. Eventually, we need to
 study how  to choose tuning parameters $\lmone$ and $\lmtwo$
 appropriately, such that the LAD-FLSA is block selection consistent.

When the true block model in \eqref{truemodel} is also sparse,
we need the following additional  conditions
to separate nonzero blocks from zero ones.
\begin{itemize}
\item[(C1):]  (a) $\lmone (b_{\min}^0)^{1/2}\to \infty$ when $n\to \infty$;
     (b) there exists $\delta>0$, such that $\lmone(b^0_{\min}/\log(J_0-K_0))^{1/2}>4\sqrt{2}(1+\delta)$.
\item [(C2):]    $\lmtwo/b_{\min}^0 <\lmone/8$ for sufficiently large $n$.
\item [(C3):]  (a) $\rho_n(b^0_{\min})^{1/2} \to \infty$ when $n\to \infty$; (b) there exists $\delta>0$
such that $\rho_n (b^0_{\min}/\log(K_0))^{1/2}>2\sqrt{2}(1+\delta)/f(0)$.
\item [(C4):]  $\lmtwo/b^0_{\min}<f(0)\rho_n/3$ for sufficiently large $n$.
\item [(C5):]  $\lmone<f(0)\rho_n/2$ for sufficiently large $n$.
\end{itemize}
Here Condition (C1) and (C2) indicate
 that  either $\lmone$ or the smallest block size $b^0_{\min}$ should
 grow with $n$ with a lower bound
 provided in (C1-b) since
  $\lmtwo$ grows with $n$ from (B1).
  Especially, if  $\lmone$
 is relatively small as seen in (C5), $b^0_{\min}$
  must be large enough.
   (C4) and (C5) provide us  a lower bound
   for the smallest nonzero signal $\rho_n$ when $n$ is  large.
    Above interpretations are consistent with (C3-a), which
    requires either the block size or the true nonzero signal intensities
 should be large enough such that the nonzero blocks can be separated from zero ones.
 In other words,  if $\rho_n$ is relatively smaller, it becomes harder to separate
 nonzero ones from zero ones. However, it is not impossible
 for us to distinguish those nonzero blocks if we have larger enough block size
   since more observations can be used to
estimate $\nu_j^0$ within $j$th block.
   (C3-b)  provides us an upper bound of the number of nonzero blocks.
  It is worthwhile to point out that even though these conditions
  seem to be complicated, some of them
  can be redundant. For instance,
  (C2) and (C5) can be used to derive a smaller upper bound than the one in (C4). Thus, if both (C2) and (C5)
 are satisfied, (C4) can be redundant.
 One can see that (C3-a) can be also redundant if (C5) and (C1-a) hold.

\begin{theorem}\label{theorem-consistency-flsa}
Under (A1), (B1-B3) and (C1-C5), a LAD-FLSA solution is  block  sign consistent.
\end{theorem}
 Theorem \ref{theorem-consistency-flsa} tells us that
 the LAD-FLSA can first recover the block patterns of hidden signals  by detecting all the true jumps,
 and then rule out those nonzero blocks.
 Furthermore, with a very large probability, those nonzero blocks
 are identified correctly
  to have either positive or negative
 signals.
Thus, the LAD-FLSA is justified to be  a promising approach for signal processing
when the true hidden signal vector  is both blocky and sparse and the observed data are contaminated by outliers.
The proof of Theorem \ref{theorem-consistency-flsa} is
provided in the Appendix.

{\it Remark 1: The block assumption of the true model in \eqref{truemodel} is  crucial in our
study. If the model is grouped, but not blocky,  fused lasso might be misleading since the
 fusion term is used to generate the block-wise solution.
 Some other techniques such as group lasso (Yuan and Lin, 2006)
 or smooth lasso (Hebiri, 2008) can be more useful to generate the
 corresponding group sparsity structure.}

{\it Remark 2: The relaxation of Gaussian or sub-Gaussian random error assumption is
important since it is very common  to see some contaminated data in signal processing, especially when
repeated measurements are not available. Some normalization methods such as Loess have been used
in  preprocessing the real data in order to improve the robustness of LS-FLSA. However, those techniques may over-smooth the data and then generate some
false negatives.}

\subsection{Additional remarks on asymptotic properties}
We will provide two additional comments on the asymptotic results obtained in
Section 3 and 4.

\noindent{\it Remark 3: An LAD-FLSA may not reach the estimation consistency and sign consistency simultaneously.}

The rate estimation consistency in Theorem \ref{est-con} holds
for $\lm_{1n}+2\lm_{2n}=O(\log (n)/n)^{1/2}$. However,
from (B1-b) and (C2), we know one of the sufficient conditions for the sign consistency in
 Theorem \ref{theorem-consistency-flsa} requiring  $\lm_{kn}>O(\log (n))^{1/2}$ for $k=1, 2$.
So an LAD-FLSA may not
be able to reach both the estimation consistency and sign consistency simultaneously.
However, this claim is not theoretically justified since
 all conditions assumed in both Theorem \ref{est-con}
and \ref{theorem-consistency-flsa} are sufficient.

\noindent {\it Remark 4: The weak irrepresentable condition is not necessary
for the jump point detection consistency in Theorem \ref{theorem-consistency-fsa}.}

To understand Remark 4, we will transform the signal approximation model in  \eqref{LAD-FSA} into a
Lasso representation.  Consider a linear regression model
\bel{lasso general}
y_i=\sum_{j=1}^p x_{ij} \beta_j +\veps_i, \quad 1\le i\le n,
\eel
where
$(y_i, x_{i1},\cdots,x_{ip})$ and
 $\bbeta=(\beta_1,\cdots,\beta_p)' $ represent the observed data and
 coefficients vector.
A  Lasso solution (Tibshirani, 1996) of $\bbeta$ is
$$
\hbbeta (\lm)=\argmin\left\{(1/2)\sum_{i=1}^n (y_i-\sum_{j=1}^p x_{ij}\beta_j)^2 +\lm \sum_{j=1}^p|\beta_j|\right\}.
$$
If we further divide the coefficients vector $\bbeta=(\bbeta_{\bone}', \bbeta_{\btwo}')'$, where
$\bbeta_{\bone}$  include those nonzero coefficients and $\bbeta_{\btwo}$ includes zeros only,
and correspondingly, we can write  $\bX=(\bX_{\bone},\bX_{\btwo})$ and
 $\bs_{\bone}=\sgn(\bbeta_{\bone})$ consist of sign mappings of  non-zero coefficients in the true model,
then  the   {\it weak irrepresentable condition}  of the designed matrix $\bX$ means
\bel{wic}
|\bX'_{\btwo}\bX_{\bone}(\bX'_{\bone}\bX_{\bone})^{-1}\bs_{\bone}|<\bone,
\eel
where $\bone$ is a vector with element being 1.
Naturally, we can write the LAD-FSA in  \eqref{LAD-FSA}  into a Lasso solution of $\hnu=(\nu_1,\cdots,\nu_n)'$,
\bel{lasso model}
  \hbnu_n^{\rm F}(\lm_{2n})=\argmin\left\{\|\by-\bZ\bnu\|_2
+\lambda_{2n}\sum_{i=2}^n|\nu_i|\right\},
\eel
 where $\nu_1=\mu_1$, $\nu_i=\mu_i-\mu_{i-1}$ for $2\le i\le n$ and
 $\bZ$ is the low triangular design matrix with nonzero items being 1.
Zhao and Yu (2006) proved that the weak irrepresentable condition
is a necessary condition for a Lasso solution in \eqref{lasso general} to be sign consistent under two regularity conditions.
We list the result in the following Lemma \ref{lem:wic}.
\begin{lemma}\label{lem:wic}
(Zhao and Yu, 2006) Suppose two regularity conditions
are satisfied for the designed matrix $\bX$:
(1) there exists a positive definite matrix $C$ such that
the covariance matrix $\bX'\bX/n\to C$ as $n\to \infty$, and
(2) $\max_{1\le i\le n}\bx_i'\bx_i/n \to 0$ as $n\to \infty.$
Then Lasso is general sign
consistent, $\lim_{n\to \infty} P(\exists \lm\ge 0, \sgn(\hbbeta(\lm))=\sgn(\bbeta_{\bzero}))=1$,
only if there exists $N$ so that $\bX$ satisfies the weak irrepresentable condition holds for $n > N$.
Here $\bbeta_{\bzero}$ is the true coefficient vector.
\end{lemma}
Unfortunately, it is easy for us to verify that
the design matrix $\bZ$ in \eqref{lasso model}
 does not satisfy the weak irrepresentable condition.
For example, if we consider a signal approximation data with only five observations where
$\mu_1\neq\mu_2\neq\mu_3=\mu_4=\mu_5$, then the first row vector of
$\bZ'_{\btwo}\bZ_{\bone}(\bZ'_{\bone}\bZ_{\bone})^{-1}$ is $(0,0,1)'$
Thus \eqref{wic} is violated.
However, there is no contradiction between the sign consistency result in Theorem  \ref{theorem-consistency-fsa}.
and Lemma \ref{lem:wic} since both two regularity conditions
in Lemma \ref{lem:wic} are violated for design matrix $\bZ$.
Suppose $\rho_1\le \cdots \rho_n$ are eigenvalues of $\bZ'\bZ/n$.
we know that (a) $\rho_1<1/(3n)\to 0$ and $\rho_n>4n^{1/2}\to \infty$ when $n\to \infty$, and in addition,
(b)  $\max_{1\le i\le n} \bz_i'\bz_i/n =1$.

\section{ Degrees of freedom of LAD-FLSA}\label{GDF}
It is crucial to seek appropriate $\lmone$ and $\lmtwo$
in (\ref{lad-flsa model}).
Large $\lmone$ will generate all zero coefficients, while large $\lmtwo$ will generate all zero jumps.
 Conditions on $\lmone$ and $\lmtwo$ in Section \ref{est-con} and \ref{sign con}
 provide us some guidance   in choosing
 the rates of two tuning parameters to obtain a
 well-behaved LAD-FLSA estimate. This section helps
 us to choose two optimal tuning parameters from the model selection point of view.

 For given $\lm_1$ and $\lm_2$, a LAD-FLSA approach
 is a modeling procedure including both model selection and model fitting.
 The complexity of a modeling procedure
 is defined  as the generalized degrees of freedom (df)
 and measured by  the sum of the sensitivity of the predicted values.
See  Ye (1998) and Gao and Fang (2011) for the discussion on the df for a modeling procedure under both the $\ell_2$
 and $\ell_1$ loss functions, respectively.
 For $1\le i\le n$, let  $\hmu_i({\by;\lm_1, \lm_2})$ be a LAD-FLSA fitted value of $y_i$ for any given  $\lm_1$ and $\lm_2$.
The degrees of freedom  of  a LAD-FLSA approach,
\bel{Def:gdf}
{\rm df}(\lm_1,\lm_2)
=\sum_{i=1}^n \partial
\rE[\hmu_i({\by;\lm_1, \lm_2})]/\partial y_i.
\eel
 In Theorem \ref{thm-unbiased-fl},
we provide an unbiased estimator of df$(\lm_1,\lm_2)$ in \eqref{Def:gdf}
  for a LAD-FLSA modeling procedure.
\begin{theorem}\label{thm-unbiased-fl}
Consider a LAD-FLSA modeling procedure defined for \eqref{signal approximation model},
\eqref{lad-flsa model} and \eqref{truemodel}.
For any fixed positive $\lm_1$ and $\lm_2$, we have
\bel{eq-df-gfl} \rE[|\widehat{\cK}(\lm_1,\lm_2)|]={\rm df}(\lambda_1, \lambda_2). \eel
\end{theorem}
  Theorem \ref{thm-unbiased-fl}
  indicates that the number of  nonzero blocks,
$|\widehat{\cK}(\lm_1,\lm_2)|$, is
an unbiased estimator of the degrees of freedom
of a LAD-FLSA modeling procedure with any given  $\lm_1$ and
$\lm_2$.
We provide both the  numerical demonstration and  theoretical proof are provided in Section \ref{sec-sim-gdf}
and the Appendix, respectively.
In fact, such an unbiased estimator  in \eqref{eq-df-gfl} can be also observed from Theorem 2 of Li and Zhu (2008).
 For example, if $\lm_1=0$,  for any $\lm_2>0$,
  the LAD-FLSA reduces to a LAD-LASSO solution of $\bw$ ($w_i=\mu_i-\mu_{i-1}$) for $2\le i\le n$.
Then
$
\sum_{i=1}^n\partial \widehat y_i(0,\lm_2)/\partial y_i=|\widehat \cJ(0,\lm_2)|.
$
Suppose  $\lm_2>0$ is fixed, the block partition is decided. Then
 for $\lm_1>0$, the LAD-FLSA becomes a LASSO model of $\bnu$ ($\bnu$ is the
block intensity vector).
Therefore,
$
\sum_{i=1}^n \partial \widehat y_i(\lm_1,\lm_2)/\partial y_i=|\widehat \cK(\lm,\lm_2)|.
$

 Results in Theorem \ref{thm-unbiased-fl}
  can be used to
choose two optimal tuning parameters
 from the model selection point of view.
 Let $y_i^0$'s denote new observations generated
from the same mechanism generating $y_i$'s.
The prediction error of $\hbmu(\by;\lambda_1, \lambda_2)$ is defined as
 \bel{loss}
\rE_0\{\sum_{i=1}^n |\hmu_i(\lambda_1, \lambda_2)-y_i^{0}|\},
 \eel
where $\rE_0$ is taken over $y_i^0$'s. From Theorem \ref{thm-unbiased-fl},
we can estimate the prediction error \eqref{loss} by
$$
\sum_{i=1}^n |y_i-\hmu_i(\lambda_1, \lambda_2)| +|\widehat{\cK}(\lm_1,\lm_2)|.
$$
Thus some existing model
selection criteria can be modified to choose an optimal
combination of tuning parameters. For instance, we can extend
 AICR (Ronchetti, 1985),  BIC (Schwarz, 1978) and GCV (Wahba, 1990) to the LAD-FLSA as follows,
\bel{eq:aicr}
\begin{array}{ll}
&{\rm AICR:}\quad \sum_{i=1}^n|y_i
-\hmu_i(\lm_1,\lm_2)|
            +|\widehat{\cK}(\lm_1,\lm_2)|,\\
&{\rm BIC:}\quad \sum_{i=1}^n|y_i -\hmu_i(\lm_1,\lm_2)|
+|\widehat{\cK}(\lm_1,\lm_2)|\log(n)/2,\\
&{\rm GCV:} \quad\sum_{i=1}^n|y_i
-\hmu_i(\lm_1,\lm_2)|/[1-|\widehat{\cK}(\lm_1,\lm_2)|/n].
\end{array}
\eel

\section{Numerical studies}

In this section, we first use some simulation
studies and real data analysis to demonstrate the performances of the LAD-FLSA approach
in recovering the true hidden signals.
 Then we verify Theorem \ref{thm-unbiased-fl} numerically
 using a sample copy number data.

\subsection{Recovery of hidden signals}\label{sec-sim-con}

 We illustrate the performance of the LAD-FLSA
 by modifying the block example studied in both Donoho and Johnstone (1995) and  Harchaoui and L\'{e}vy-Leduc (2010),
 where the signal vector is only blocky but not sparse.
We choose ${\bf t}=(.1, .23, .65, .76, .9)'$ and ${\bf h}=(1.5,    -3,   4.3, -3.1,  -2)'$
and round $\sum_j h_j(1 + \sgn (i/n-t_j))/2$
to the nearest integers to get $\mu_i^0$ for $1\le i\le n$.
Then the generated true hidden signal vector,
$$\bmu^0=(\bzero_{p_1}' \btwo_{p_2}' -\btwo_{p_3}' {\bf 3}_{p_4}' \bzero_{p_5}' \btwo'_{n-q})'$$
is blocky and sparse with four nonzero blocks and two zero ones. Here $q=p_1+\cdots+ p_5$.
The observed data are generated from model \eqref{signal approximation model} by
simulating
$\veps_i$'s from 1) normal distribution with mean $0$ and standard deviation $\sigma$,
2) double exponential distribution with center $0$
and standard deviation $\sigma$, 
and 3) standard cauchy distribution
with a multiplier $0.1\sigma$.
 Similar to  Harchaoui and L\'{e}vy-Leduc (2010), we consider
weak, mild and strong noises by setting $\sigma=0.1$, $0.5$ and $1$ in all
three types of distributions.
In Figure \ref{fig:signaldata}, we plot
a sample data set generated from 2),
where the observed data, the true hidden signals and the
LAD-FLSA estimates are plotted using gray, black and red colors, respectively.

The data is standardized as $y_i/s(\by)$ and analyzed  using
the LAD-FLSA approach in \eqref{lad-flsa model}, where $s(\by)$ is
the standard deviation of $(y_1,\cdots,y_n)'$.
We choose ``optimal'' $\lm_1$ and $\lm_2$ by minimizing  BIC in \eqref{eq:aicr}
 for $0<\lm_1<0.5$  with
increments of $0.01$ and  $(n/\log(n))^{1/2}<\lm_{2n}<n^{1/2}$ with
increments of $0.1$, respectively.
To demonstrate the robust properties  of the LAD-FLSA approach,
we also report the simulation results from
the LS-FLSA  approach in \eqref{tw-ls flsa model}.
For each model, we illustrate the variable selection effect using CFR+6,
 the ratio of either recovering $\hmu^0$ correctly or
 over-fitting the model by including six additional noises
  over $1000$ replicates.
  We choose ``six'' here since we have six blocks in
the true model.
We also report JUMP, the average number (with standard deviation) of
jumps over $1000$ replicates. A jump is counted only if
the adjacent differences is at least $0.1$.
We demonstrate the estimation effects by computing the least absolute relative error (LARE)
as follows:
\bel{eq:lare}
{\rm LARE}(\hbmu_n,\bmu^0)=\dfrac{\sum_{i=1}^n|\hmu_i-\mu_i^0|}{\sum_{i=1}^n|\mu_i|}.
\eel
The simulation results for sample size $n=1000$ and $5000$ are reported
in Table \ref{sim results}, where we can see that
the LAD-FLSA approach has much better performance than the
LS-FLSA for both strong and mild signal noises.
When the signal noises are weak, the LAD-FLSA still has some advantages over
the LS-FLSA especially when the data is contaminated by Cauchy distributed noises. For example,
for Cauchy error and $\sigma=0.1$,
the LAD-FLSA recovers the true signal vector exactly at a ratio of 87\% for $n=1000$ and  92\%
for $n=5000$, while
the LS-FLSA only recovers the true model exactly 49\% and 78\% of the time.

\subsection{BAC array}\label{sec-bac}

In Section 1 we introduced a sample BAC CGH data, where
the observation of each entry for cell line GM 13330 is the log $2$
fluorescence ratios from all $23$ chromosomes resulted from the BAC experiment sorted in the
order of the clones locations on the genome. The purpose of
the study is to detect the locations where there are some significant
deletions or amplifications. As a demonstration of  the effect of the LAD-FLSA
applied to copy number analysis, we only analyze the data
from chromosome 1--4 with $129$, $67$, $83$ and $167$ markers, respectively.
  Since the log $2$ ratios at many markers are observed to be around $0$
  and the data may also have some spatial
dependence properties, it is reasonable to
assume the true hidden signals to be both  sparse and blocky.
We analyze each chromosome independently by using both
the LAD-FLSA in \eqref{lad-flsa model} and LS-FLSA
in \eqref{tw-ls flsa model}.
 Tuning parameters are chosen the same as in Section \ref{sec-sim-con}.
The final estimates from both methods on all $4$ chromosomes are plotted together in Figure 1.
The LAD-FLSA estimates (top panel) provides four blocks with one amplification region
in chromosome 1 and one deletion region in chromosome 4.
Besides the two variation regions detected by the LAD-FLSA,
the LS-FLSA estimates (bottom panel) also show an amplification at
a single point in chromosome 2, which is not confirmed by spectral karyotyping
in  Snijders et al. (2001).

 \subsection{Effect of unbiased estimator of the degrees of freedom}\label{sec-sim-gdf}

We now conduct some simulations based upon the sample BAC array
studied in Section \ref{sec-bac} to examine
 Theorem \ref{thm-unbiased-fl} numerically.
To illustrate the effect of the unbiased estimator of the degrees of freedom, we only take
chromosome $1$ with  $129$ locations as an example. One can
see the sample data from Figure 1.

We generate $500$ Monte Carlo simulations based on the
same hypothetical model
$$y_i^0=y_i+\veps^0_i,
~~i=1, \cdots, 129,$$ where $y_i$s are observations at $129$ locations and
 $\veps^0_i$'s are independent normal
 with center $0$ and  standard deviation $0.1\sigma^*$, where
$\sigma^*$ is the standard deviation of $\by$.
For each combined $(\lm_1,
\lm_2)$ with $0<\lm_1,\lm_2\le 1$, we record
$\widehat{\mbox{df}}(\lm_1,\lm_2)$ from $|\widehat{\cK}(\lm_1,\lm_2)|$,
and  compute the true
$\mbox{df}(\lm_1,\lm_2)$
defined in \eqref{Def:gdf} using the Monte Carlo simulation from Algorithm 1 in Ye
(1998). In Figure 2, we
plot
$\widehat{\mbox{df}}(\lm_1,\lm_2)$ of the LAD-FLSA estimate
 for  every combination of $0<\lm_1,\lm_2\le 1$,
 with the increment of $0.05$, respectively.
The averages of df$(\lm_1,\lm_2)$ and
$|\widehat \cK(\lm_1,\lm_2)|$ over 500 repetitions are reported
  in Figure
\ref{fig:CGHdf}. Those simulation results show that
the number of estimated nonzero blocks $|\widehat \cK(\lm_1,\lm_2)|$ is a promising estimate to
the df$(\lm_1,\lm_2)$ numerically, especially when the
number of estimated nonzero blocks is not  deviated from the true one seriously.

\section{Concluding remarks}

In this paper, we study the asymptotic properties of the LAD signal-approximation approach
using the fused lasso penalty.
By assuming the true model to be both blocky and sparse,
we investigate both the estimation consistency and sign consistency  of the LAD-FLSA  estimator.
In terms of estimation consistency, the consistency rate
is optimal up to an logarithmic factor  if the dimension of any linear space where the true model and its estimates belong is bounded from above.
In terms of sign consistency,  we justify that
a LAD-FLSA approach can not only recover the true block pattern
but also distinguish those nonzero blocks from the zero ones
 correctly with high probability under reasonable conditions.
In fact, those jump selection and block selection consistency
results can be made stronger by matching the
corresponding signs correctly with a large probability.
Thus, by choosing two tuning
parameters $\lm_1$ and $\lm_2$ properly, we can reach a well-behaved
LAD-FLSA estimate to recover the true hidden signal vector under some random noises.
The consistency results in this paper extend
 the theoretical properties of the LS-FSA in  Harchaoui and L\'{e}vy-Leduc (2010) and
  the LS-FLSA in Rinaldo (2009) to the LAD signal approximation, which
 amplify the study of signal approximation using
linear regression when the random error does not
follow a Gaussian distribution. Furthermore,
we demonstrate that the number of estimated nonzero blocks
is an unbiased estimator of the degrees of freedom
 of the LAD-FLSA.
Thus, the existing  model selection criteria
can be extended to the LAD-FLSA for choosing the tuning parameters.

As in many recent studies, our results are proved for penalty parameters that satisfy the conditions as stated in the theorems. It is not clear whether the penalty parameters selected using data-driven procedures satisfy those conditions. However, our numerical study shows a satisfactory finite-sample performance of the LAD-FLSA. Particularly, we note that the tuning parameters selected based on the BIC seem sufficient for our simulated data. This is an important and challenging problem that requires further investigation, but is beyond the scope of the current paper.
Also, a basic assumption required in our results is that the random error terms $\varepsilon_i$ in  (\ref{signal approximation model}) are independent. Since the observations $y_1, \ldots, y_n$ are in a natural order in this model, for example, copy number variation data based on genetic markers are ordered according to their chromosomal locations, it would be interesting to study the behavior of the LAD-FLSA allowing for certain dependence structure in the error terms.


\section *{Appendix}

\noindent{\bf Proof of Lemma \ref{lemma dmax}}

 \noindent
Let $\widetilde w_i=\widetilde w_i(\lmone,\lmtwo)=\tmu_i-\tmu_{i-1}$
and $\widehat w_i=\widehat w_i(\lmone,\lmtwo)=\hmu_i-\hmu_{i-1}$  be
the LS-FLSA and LAD-FLSA estimates of $i$th jump in
\eqref{ls-flsa model} and \eqref{lad-flsa model}.
 Using the Karush--Kuhn--Tucker (KKT) conditions \eqref{ls-flsa model}, we get
 \bes
  -2\sqrt{f(0)} (z_i-\sqrt{f(0)} \sum_{j=1}^i\widetilde w_j)+\lmone \sum_{k=i}^n \sgn(\sum_{j=1}^k \widetilde w_j)=-\lmtwo \sgn(\widetilde w_i)  &{\rm if}~\widetilde w_i\neq 0.
  \ees
Then
\bes
\lmtwo^2\widetilde {\cJ}\le 8f(0)\sum_{i=1}^n(z_i-\sqrt{f(0)}\sum_{j=1}^i\widetilde w_j)^2+2\lmone^2 n^2 \widetilde {\cJ}.
 \ees
Thus
\bel{cj1}|\widetilde {\cJ}|\le  8f(0)(\lmtwo^2-2 n^2\lmone^2)^{-1}\sum_{i=1}^n z_i^2\le 16f(0)n/(\lmtwo^2-2 n^2\lmone^2).\eel
Using the  KKT equations of  \eqref{lad-flsa model}, we have
  \bes
  -\sgn (y_i-\sum_{j=1}^i\widehat w_j)+\lmone \sum_{k=i}^n \sgn(\sum_{j=1}^k \widehat w_j)=-\lmtwo \sgn(\widehat w_i)  &{\rm if}~\widehat w_i\neq 0.
  \ees
Then
\bel{cj2}
|\widehat {\cJ}|\le n/(\lmtwo-\lmone n).
 \eel
Finally, combining with \eqref{cj1}, \eqref{cj2} and (A2), (iii) holds,
which completes the proof of Lemma \ref{lemma dmax}.
$\Box$

 \vspace{10pt}

 \noindent{\bf Proof of Lemma \ref{ls-flsa consistency} }

 \noindent
    From the definition of $\tbmu_n$ in \eqref{ls-flsa model}, we have
 \bel{eq-con-1}
\begin{array}{ll}
 \sum_{i=1}^n(z_i-\sqrt{f(0)}\mu_i)^2&\le \sum_{i=1}^n(z_i-\sqrt{f(0)}\mu_i^0)^2 \\
  &+\lmone  \sum_{i=1}^n[|\mu_i^0|-|\tmu_i|]
  +\lmtwo\sum_{i=2}^n[|\mu_i^0-\mu_{i-1}^0|-|\tmu_i-\tmu_{i-1}|.
  \end{array}
 \eel
 From the triangle inequality,
  \eqref{eq-con-1} becomes
 \bel{eq-con-2}
 \begin{array}{ll}
 &f(0)\sum_{i=1}^n (\tmu_i-\mu^0_i)^2 \\
 &\quad\le 2\sqrt{f(0)} \sum_{i=1}^n \eta_i(\tmu_i-\mu_i^0)+
 \lmone\sum_{i=1}^n|\tmu_i-\mu_i^0|+\lmtwo\sum_{i=2}^n[|\mu_i^0-\mu_{i-1}^0|-|\tmu_i-\tmu_{i-1}|]\\
  &\quad\le 2\sqrt{f(0)} \sum_{i=1}^n \eta_i(\tmu_i-\mu_i^0)+
 (\lmone+2\lmtwo)\sum_{i=1}^n|\tmu_i-\mu_i^0|.
 \end{array}
 \eel
 The rest of the proof is similar to the proof of Proposition 2 in Harchaoui and L\'{e}vy-Leduc (2010).
 For $\bmu\in \cR^n$, we define $$G(\bmu)=2\sqrt{f(0)}\sum_{i=1}^n\eta_i(\mu_i-\mu_i^0)/\|\bmu-\bmu^0\|_2.$$
  Thus \eqref{eq-con-2} becomes
 $$
f(0)\sum_{i=1}^n(\tmu_i-\mu_i^0)^2\\
   \le  (\lmone+2\lmtwo) \sqrt{n}\|\tbmu_{n}-\bmu^0\|_2 +G(\tbmu_{n})\|\tbmu_{n}-\bmu^0\|_2.
  $$
 Then,
 \bel{eq-con-3}
 \sqrt{f(0)}\|\tbmu_{n}-\bmu^0\|_2\le  (\lmone+2\lmtwo)  \sqrt{n}+ G(\tbmu_{n}).
 \eel
Let $\{S_K\}$ be a collection of any $K$-dimensional linear space to which $\tbmu_n$ may belong.
 From Lemma \ref{lemma dmax}, $1\le K\le \Lambda_n$.
From \eqref{eq-con-3}, for any $\delta_n>0$,
 \bel{eq-con-4}
 \begin{array}{ll}
 \bP(\|\tbmu_n-\bmu^0\|_2\ge \delta_n)
 &\le \bP(G(\tbmu_{n})\ge \sqrt{f(0)}\delta_n-(\lmone+2\lmtwo) \sqrt{n} )\\
 & \le \sum_{k=1}^{\Lambda_n} n^k \bP\left(\sup_{\bmu\in S_K} G(\bmu)\ge \sqrt{f(0)}\delta_n-(\lmone+2\lmtwo) \sqrt{n} \right).
 \end{array}
 \eel
 Notice that $\bE(G(\bmu))=0$ and $\Var(G(\bmu))=1$. As a consequence of
Cirel'son, Ibragimov and Sudakov's (1976) inequality,
 \bel{eq-con-5}
 \bP\{\sup_{\bmu \in S_K} G(\bmu)\ge \bE\left[\sup_{\bmu \in S_K} G(\bmu)\right]+z\} \le \exp\{-z^2/2\}
  {\rm ~for~ some~  constant~} z>0.
 \eel
Consider the collection $\{S_K\}$. Let  $\Omega$ be the $D$-dimensional space
to which $\bmu-\bmu^0$ belongs and $\bpsi_1,\cdots,\bpsi_D$
be its orthogonal basis.
\bel{eq-con-6}
\begin{array}{ll}
\sup_{\bmu \in S_K} G(\bmu)&\le \sup_{\omega \in \Omega}\dfrac{2\sqrt{f(0)}\sum_{i=1}^n \eta_i \omega_i}{\sqrt{n}\|\omega\|_n}\\
&=\sup_{\bf a \in \cR^D}\dfrac{2\sqrt{f(0)}\sum_{i=1}^n \eta_i (\sum_{j=1}^D a_j\psi_{j,i})}
                            {\sqrt{n}\|\sum_{j=1}^D a_j {\bpsi_j}\|_n}\\
&=\sup_{\bf a \in \cR^D}\dfrac{2\sqrt{f(0)}\sum_{j=1}^D a_j (\sum_{i=1}^n \eta_i\psi_{j,i})}
                            {(\sum_{j=1}^D a_j)^{1/2}}\\
&\le 2\sqrt{f(0)}\left(\sum_{j=1}^D (\sum_{i=1}^n \eta_i\psi_{j,i})^2 \right)^{1/2}),
\end{array}
\eel
where the last ``$\le$'' is obtained using the Cauchy-Schwarz inequality.
From (A2) and (i) in Lemma \ref{lemma dmax},
there exists $M_1>0$ such that
$D< (M_1+1)\Lambda_n$. Then by taking expectations on both sides of \eqref{eq-con-6}, we have
\bel{eq-con-7}
\begin{array}{ll}
\bE[\sup_{\bmu \in S_K} G(\bmu)]
 &\le 2\sqrt{f(0)} \bE\left [\left (\sum_{j=1}^D \left (\sum_{i=1}^n \eta_i\psi_{j,i}\right )^2 \right )^{1/2}\right ] \\
& \le   2\sqrt{f(0)} \left ( \sum_{j=1}^D \bE \left[  \left(\sum_{i=1}^n \eta_i\psi_{j,i}\right )^2 \right] \right )^{1/2}\\
&\le \sqrt{D} \le ((M_1+1)\Lambda_n)^{1/2}.
\end{array}
\eel
Combining \eqref{eq-con-5} and \eqref{eq-con-7}, we get
\bel{eq-con-8}
\bP\left(\sup_{\bmu \in S_K} G(\bmu)\ge ((M_1+1)\Lambda_n)^{1/2}+z\right) \le \exp\{-z^2/2\}.
\eel
Let $0<c<1$ such that
$$
c\sqrt{f(0)}\delta_n=(\lmone+2\lmtwo)\sqrt{n}+((M_1+1)\Lambda_n)^{1/2}.
$$
Then we can choose a positive
$z=\sqrt{f(0)}\delta_n-(\lmone+2\lmtwo)\sqrt{n}-((M_1+1)\Lambda_n)^{1/2}$ in \eqref{eq-con-8}.
Combining \eqref{eq-con-4} and \eqref{eq-con-8},
\bel{eq-con-9}
\begin{array}{ll}
\bP(\|\tbmu_{n}-\bmu^0\|_2\ge \delta_n) &\le \Lambda_n\exp\{\Lambda_n \log n -(1/2)[\sqrt{f(0)}\delta_n-(\lmone+2\lmtwo)\sqrt{n}-(M_1+1)\Lambda_n]^2\}\\
&\le \Lambda_n\exp\{\Lambda_n \log n -(1/2)(1-c)^2f(0)\delta_n^2\}.
\end{array}
\eel
For $\alpha_n=\delta_n/\sqrt{n}$, we have
$$
\bP(\|\tbmu_n-\bmu^0\|_n \ge \alpha_n)
\le \Lambda_n \exp\{\Lambda_n \log n -(1/2)(1-c)^2f(0) n\alpha_n^2\}.
$$
Thus the first part of Lemma~\ref{ls-flsa consistency} holds.
Furthermore, if we also have $\alpha_n=\{2M_2 \Lambda_n (\log n)/ n\}^{1/2}$, then
$$
\bP(\|\tbmu_n-\bmu^0\|_n \ge \sqrt {2M_2 \Lambda_n (\log n)/ n})
\le \Lambda_n n^{\{1-M_2f(0)(1-c)^2\}\Lambda_n},
$$ which completes the proof.
 $\Box$

\vspace{15pt}

 \noindent {\bf Proof of Theorem \ref{lad-flsa consistency}}

 \noindent
   Define
   $$
   L_n(\bmu)=n^{-1}\left[\sum_{i=1}^n|y_i-\mu_i|-\sum_{i=1}^n|y_i-\mu^0_i|
       +\lmone\sum_{i=1}^n|\mu_i|+\lmtwo\sum_{i=2}^n|\mu_i-\mu_{i-1}|\right]
   $$
   and
    $$
   M_n(\bmu)=n^{-1}\left[f(0)\sum_{i=1}^n(\mu_i-\mu_i^0)^2-\sum_{i=1}^n\sgn(\veps_i)(\mu_i-\mu_i^0)
       +\lmone\sum_{i=1}^n|\mu_i|+\lmtwo\sum_{i=2}^n|\mu_i-\mu_{i-1}|\right].
   $$
   Then $\hbmu_n=\argmin\{L_n(\bmu)\}$ and $\tbmu_n=\argmin\{M_n(\bmu)\}$.
   Define $R_{ni}=R_{ni}(\mu_i,\veps_i)=|\veps_i-(\mu_i-\mu_i^0)|-|\veps_i|+\sgn(\veps_i)(\mu_i-\mu_i^0)$ and $\xi_{ni}=R_{ni}-\bE[R_{ni}]$. Following Gao and Huang (2010b), we can verify
  \bel{eq-con-lad-1}
   |L_n(\bmu)-M_n(\bmu)|=\allsum\xi_{ni}/n+\tau_n (\bmu,\bmu^0),
  \eel
  where $\tau_n (\bmu,\bmu^0)=o(\|\bmu-\bmu^0\|_n^2)$.
   For any $\delta>0$, we define $S_{\delta}=\{\bmu:\|\bmu-\tbmu_n\|_2 \le \delta\}$,
   $S^d_{\delta}=\{\bmu:\|\bmu-\tbmu_n\|_2 = \delta\}$ and
   $S^0_{\delta}=\{\bmu:\|\bmu-\bmu^0\|_2 \le \delta\}$.
   We define
   $$
   \Delta_n(\delta)=\sup_{\bmu\in S_{\delta}}|L_n(\bmu)-M_n(\bmu)|
      $$
   and
   $$
   h_n(\delta)=\inf_{\bmu\in S^d_{\delta}}(M_n(\bmu)-M_n(\tbmu_n)).
   $$
   We have
   \bel{eq-con-lad-2}
   \begin{array}{ll}
   M_n(\bmu)-M_n(\tbmu_n)&=f(0)\|\bmu-\tbmu_n\|_n^2+2n^{-1}f(0)\allsum(\mu_i-\tmu_i)(\tmu_i-\mu_i^0)
    +n^{-1}\allsum\sgn(\veps_i)(\tmu_i-\mu_i)\\
    &+n^{-1}\lmone\allsum[|\mu_i|-|\tmu_i|]+n^{-1}\lmtwo\sum_{i=2}^n[|\mu_i-\mu_{i-1}|-|\tmu_i-\tmu_{i-1}|].
   \end{array}
   \eel
   Since $\partial M_n(\bmu)/\partial \mu_i\vert_{\mu_i=\tmu_i}=0$,
   $$
   2n^{-1}f(0)(\tmu_i-\mu_i^0)-n^{-1}\sgn(\veps_i)+n^{-1}\lmtwo[\sgn(\tmu_i-\mu_{i-1})-\sgn(\mu_{i+1}-\tmu_i)]=0.
   $$
  Multiply $(\mu_i-\tmu_i)$ on both sides and take sums.
  Then \eqref{eq-con-lad-2} becomes
    $$
   \begin{array}{ll}
   M_n(\bmu)-M_n(\tbmu_n)
   &=f(0)\|\bmu-\tbmu_n\|_n^2 +n^{-1}\lmone\allsum[|\tmu_i|-\mu_i\sgn(\tmu_i)-(|\tmu_i|-|\mu_i|)]\\
   &\quad +n^{-1}\lmtwo\sum_{i=2}^n[ \sgn(\mu_{i+1}-\tmu_i)(\mu_i-\tmu_i)
           + \sgn(\tmu_{i}-\mu_{i-1})(\tmu_i-\mu_i)]\\
         & \quad +   n^{-1} \lmtwo\sum_{i=2}^n[|\mu_i-\mu_{i-1}|-|\tmu_i-\tmu_{i-1}|]\\
    &>f(0)\|\bmu-\tbmu_n\|_n^2\\
             &\quad +n^{-1}\lmtwo\sum_{i=2}^n[ \sgn(\mu_{i+1}-\tmu_i)(\mu_i-\tmu_i)
           + \sgn(\tmu_{i}-\mu_{i-1})(\tmu_i-\mu_i)]\\
         & \quad +   n^{-1} \lmtwo\sum_{i=2}^n[|\mu_i-\mu_{i-1}|-|\tmu_i-\tmu_{i-1}|].
    \end{array}
$$
  Then for any $\kappa_{n}>0$, we have $$ h_n(\kappa_{n})>f(0)\kappa_{n}^2/n.$$
  From the Convex Minimization Theorem in Hjort and Pollard (1993), we have
  \bel{eq-con-lad-2}
  \begin{array}{ll}
  & \bP(\|\hbmu_n-\tbmu_n\|_2 \ge \kappa_{n}) \\
  &\quad\le \bP(\Delta_n(\kappa_{n})>h_n(\kappa_{n})/2) \\
  &\quad\le \bP\left(\sup_{\bmu\in S_{\kappa_{n}}}n^{-1}|\allsum \xi_{ni}|\ge f(0)\kappa_{n}^2/(2n)\right)
  +\bP\left(\sup_{\bmu\in S_{\kappa_{n}}}|\tau_n(\bmu,\bmu^0)|\ge f(0)\kappa_{n}^2/(2n) \right).
  \end{array}
  \eel
  Suppose $r_n=o(1)$ and $\tau_n=\tau_n(\tbmu_n,\bmu^0)=r_n\|\tbmu_n-\bmu^0\|_n^2$. Then
   \bel{eq-con-lad-3}
   \begin{array}{ll}
   &\lim_{n\to \infty}\bP\left(\sup_{\bmu\in S_{\kappa_{n}}}|\tau_n(\bmu,\bmu^0)|\ge f(0)\kappa_{n}^2/(2n) \right) \\
   &\quad\le
 \lim_{n\to \infty} \bP\left(\sup_{\bmu\in S_{\kappa_{n}}}2|r_n|\|\tbmu_n-\bmu^0\|_n^2\ge f(0)\kappa_{n}^2/(4n) \right) \\
  &\quad\quad+\lim_{n\to \infty}\bP\left(\sup_{\bmu\in S_{\kappa_{n}}}2|r_n|\|\bmu-\tbmu_n\|_n^2\ge f(0)\kappa_{n}^2/(4n) \right)  \\
 & =\lim_{n\to \infty} \bP\left(\sup_{\bmu\in S_{\kappa_{n}}}|r_n|\|\tbmu_n-\bmu^0\|_n^2\ge f(0)\kappa_{n}^2/(8n) \right).
 \end{array}
  \eel
  Combining \eqref{eq-con-lad-1} and \eqref{eq-con-lad-2},
   \bel{eq-con-lad-3}
  \begin{array}{ll}
    \bP(\|\hbmu-\tbmu\|_2 \ge \kappa_{n})
     &\le \bP\left(\sup_{\bmu\in S_{\kappa_{n}}}\sum_{i=1}^n\xi_{ni}/n >f(0)\kappa_{n}^2/(2n) \right)\\
  &\quad + \bP\left(\sup_{\bmu\in S_{\kappa_{n}}}|r_n|\|\tbmu_{n}-\bmu^0\|_n^2\ge f(0)\kappa_{n}^2/(8n) \right)+o(1)\\
  &\le \bP\left(\sup_{\bmu\in S^0_{\kappa_{n}'}}|\sum_{i=1}^n\xi_{ni}/n| >f(0)\kappa_{n}^2/(2n) \right)
  + \bP\left(\|\tbmu_{n}-\bmu^0\|_2\ge \kappa_{n}\right),
  \end{array}
  \eel
  where $\kappa_{n}'=2\kappa_{n}$.
  Let the $u_i$'s be a Rademacher sequence
  and  $\bpsi_1,\cdots,\bpsi_D$ be the orthogonal basis
  of a  $D$-dimensional space to which $\bmu-\bmu^0$ belongs.
  Using the Contraction Theorem in Ledoux and Talagrand (1991) and also
  the Cauchy-Schwarz inequality, we have
  $$
\begin{array}{ll}
  \bE\left[\sup_{\bmu\in S^0_{\kappa_{n}'}}|\sum_{i=1}^n\xi_{ni}/n|\right]
  &=(8/n)\bE\left[\sup_{\bmu\in S^0_{\kappa_{n}'}}|\sum_{i=1}^n u_i(\mu_i-\mu_i^0)|\right]\\
   &\le(8/n)\bE\left[\sup_{{\bf a}\in \cR^D}\sup_{\bmu\in S^0_{\kappa_{n}'}}|\sum_{j=1}^D a_j\sum_{i=1}^n u_i(\psi_{j,i})|\right]\\
   &\le(8/n)\bE\left[\sup_{{\bf a}\in \cR^D}\sup_{\bmu\in S^0_{\kappa_{n}'}}(\sum_{j=1}^D a_j^2)^{1/2} (\sum_{j=1}^D  (\sum_{i=1}^n u_i\psi_{j,i})^2)^{1/2}\right]\\
  &\le  (8/n) \sup_{\bmu\in S^0_{\kappa_{n}'}}\sqrt{D} (\sum_{j=1}^D a_j^2)^{1/2}\\
  &=  (8/n) \sup_{\bmu\in S^0_{\kappa_{n}'}}\sqrt{D}\|\bmu-\bmu^0\|_2\\
  &\le 16\sqrt{\Lambda_n}\kappa_{n}/n.
  \end{array}
 $$
  \bel{eq-con-lad-4}
    \begin{array}{ll}
  \bP\left(\sup_{\bmu\in S^0_{\kappa_{n}'}}|\sum_{i=1}^n\xi_{ni}/n| >f(0)\kappa_{n}^2/(2n) \right)
  &\le \bE\left[\sup_{\bmu\in S^0_{\kappa_{n}'}}|\sum_{i=1}^n\xi_{ni}/n|\right]/ (f(0)\kappa_{n}^2/(2n) )\\
  &\le  32\sqrt{\Lambda_n} / (f(0)\kappa_{n})\le 8\sqrt{\Lambda_n} / (f(0)\kappa_{n})
  .
  \end{array}
  \eel
  Let $\gamma_n=\kappa_{n}/\sqrt{n}$. Combining \eqref{eq-con-lad-3} and \eqref{eq-con-lad-4},
  $$
  \begin{array}{ll}
  \bP(\|\bmu-\bmu^0\|_n \ge \gamma_n)
 & \le \bP(\|\hbmu_n-\tbmu_n\|_2 \ge \kappa_{n}/2) +
  \bP(\|\tbmu_n-\bmu^0\|_2 \ge \kappa_{n}/2) \\
  & = \bP(\|\hbmu_{n}-\tbmu_{n}\|_2 \ge \kappa_{n}/2) +\bP(\|\tbmu_n-\bmu^0\|_2 \ge \kappa_{n}/2) \\
  &\le  32\sqrt{\Lambda_n} / (f(0)\gamma_n \sqrt{n})+  2\bP\left(\|\tbmu_{n}-\bmu^0\|_n\ge \gamma_n/2\right)\\
  &\le  32\sqrt{ \Lambda_n} / (f(0)\gamma_n \sqrt{n})+  2\Lambda_n\exp\{\Lambda_n\log n-(1-c)^2f(0)n\gamma_n^2/8\}.
  \end{array}
  $$
  The last ``$\le$'' is from \eqref{eq-con-lad-4} and Lemma \ref{ls-flsa consistency}
  by choosing $\gamma_n=2(c\sqrt{f(0)})^{-1}[\lmone+2\lmtwo+((M_1+1)\Lambda_n/n)^{1/2}]$.
   Thus the first part of Theorem \ref{ls-flsa consistency} holds.
   Furthermore, if we let
   $\lmone+2\lmtwo=[2c^2f(0)M_3\Lambda_n(\log n)/n]^{1/2}-[(M_1+1)\Lambda_n]^{1/2}$
   for $M_3>1/((1-c)^2f(0))$ and
    $\gamma_n=(8M_3\Lambda_n(\log n)/n)^{1/2}$,
    then $ \sqrt{\Lambda_n} / (f(0)\gamma_n \sqrt{n})=O(1/\sqrt{\log n})$. Thus,
    $$
    \bP(\|\hbmu_n-\bmu^0\|_n \ge \gamma_n)\le O(1/\sqrt{\log n})+2\Lambda_n\exp\{\Lambda_n\log n(1-M_3(1-c)^2f(0))\}.
    $$
$\Box$

\vspace{15pt}

\noindent{\bf Proof of Corollary \ref{corollary con}}

 \noindent
If we replace the upper bound of maximal dimension
of any linear space where $\hbmu_n$, $\tbmu_n$ or
$\bmu^0$ belong by $J_{\max}$ in the proof of Theorem \ref{lad-flsa consistency},
we can obtain the consistency result in Corollary \ref{corollary con}.
We do not repeat the proof here. $\Box$

\vspace{10pt}

\noindent{\bf Proof of Theorem \ref{theorem-consistency-fsa}}

\noindent Suppose vector $\bmu$ has $J$ blocks
and $\{\cB_1,\cdots,\cB_J\}$ is the corresponding unique block partition.
Let $\nu_j$ be  the intensity at $j$th block for $1\le j\le J$.
From Lemma A.1 in Rinaldo (2009),  the subdifferential
of the total variation penalty
\bel{subdifferential} \partial \left(\lmtwo \sum_{j=2}^J |\nu_j-\nu_{j-1}|\right)=
  \left\{
  \begin{array}{ll}
  -\lmtwo \sgn(\nu_{j+1}-\nu_j),& j=1 \\
  \lmtwo (\sgn(\nu_{j+1}-\nu_j)-\sgn(\nu_{j}-\nu_{j-1})),& 1<j<J \\
  \lmtwo \sgn(\nu_{j}-\nu_{j-1}),& j=J
  \end{array}
  \right.,
  \eel
  where $\sgn(x)=1, 0, -1$ when $x>0, =0, <0$, respectively.
 We define $c_j^0$ and $\widehat c_j$ as the subdifferentials \eqref{subdifferential}
  at both $\bnu^0$ and $\hbnu_n$ scaled by the corresponding block sizes. In other words,
 we have
  \bel{c0}
  c^0_j=\left\{
  \begin{array}{ll}
  -\lmtwo \sgn(\nu^0_{j+1}-\nu^0_j)/b^0_j,& j=1 \\
  \lmtwo (\sgn(\nu^0_{j+1}-\nu^0_j)-\sgn(\nu^0_{j}-\nu^0_{j-1}))/ b^0_j,& 1<j<J_0 \\
  \lmtwo \sgn(\nu^0_{j}-\nu^0_{j-1})/ b^0_j,& j=J_0
  \end{array}
  \right.
  \eel
    and
    \bel{chat}
    \widehat c_j=\left\{
  \begin{array}{ll}
  -\lmtwo \sgn(\hnu_{j+1}-\hnu_j)/\widehat b_j,& j=1 \\
  \lmtwo (\sgn(\hnu_{j+1}-\hnu_j)-\sgn(\hnu_{j}-\hnu_{j-1}))/\widehat b_j,& 1<j<\Jhat \\
  \lmtwo \sgn(\hnu_{j}-\hnu_{j-1})/\widehat b_j,& j=\Jhat
  \end{array}
  \right..
  \eel
 For an estimate $\hbmu_n$, we let $\widehat\cB_{j(i)}$ be the block estimate
  where $i$ stays, that is, $\hmu_i$ are all the same for $i\in \widehat\cB_{j(i)}$.
  Let $\widehat b_{j(i)}=|\widehat\cB_{j(i)}|$ be the size of  $\widehat\cB_{j(i)}$.
 Then $\cB^0_{j(i)}$ ($b^0_{j(i)}$) is the corresponding block set (size).
  From notations (I) and (V) in Section 2,
  we have $b^0_{j(i)}=|\cB^0_{j(i)}|$ for $1\le j\le J_0$.  
     From the KKT conditions, $\widehat {\bmu}^F$
   is a LAD-FSA solution  if and only if 
    \bel{kkt}
     \left\{
    \begin{array}{ll}
      \sum_{k\in \widehat \cB_{j(i)}} \sgn(y_k- \hmu_i)=  \widehat b_{j(i)}  \widehat c_{j(i)} &\quad{\rm if}~i\in \widehat\cJ \\
         |\sum_{k\in \widehat\cB_{j(i)}} \sgn(y_k-\hmu_i)|<2\lmtwo   & \quad {\rm if}~i\notin \widehat\cJ
    \end{array}
     \right..
     \eel
 Let $\hmu_i$ and $\mu_i^0$ satisfy
      \begin{equation}  \label{hmuj}
     \left\{
     \begin{array}{ll}
       \hmu_i=\mu_i^0+(2f(0)b_{j(i)}^0)^{-1}\left(\sum_{k\in \cB_{j(i)}^0} \nolimits \sgn(\veps_k)- b^0_{j(i)} c^0_{j(i)}+ \widehat h_i\right)
       &\quad \forall i\in \cJ^0\\
         \hmu_i=\hmu_{i-1} &\quad\forall i\notin \cJ^0.
     \end{array}
     \right..
    \end{equation}
  Here $\widehat h_i$ is the remainder term with the stochastically equicontinuity, more specifically,
\bel{equicontinuity}
|(b_{j(i)}^0)^{-1/2}\widehat h_{i}|=O_p(1), ~\forall {1\le i\le  n}.
\eel
In fact, $\widehat h_i=2f(0)b^0_{j(i)}  (\hmu_i-\mu_i^0)+ \sum_{k\in \cB^0_{j(i)}}\bE[r_k^i] +\sum_{ k\in \cB^0_{j(i)}}\varsigma_k^i$
 with $r_k^i=\sgn(\veps_k-(\hmu_i-\mu_i^0) )-\sgn(\veps_k)$ and $\varsigma_k^i=r_k^i-\bE_{\bveps}[r_k^i]$ for $k\in \cB^0_{j(i)}$.
    Define the difference vector $\bw=(w_1,\cdots, w_n)'$, with $w_1=\mu_1$ and $w_i=\mu_i -\mu_{i-1}$ for $2\le i\le n$.
   If $\sgn(\widehat w_i)=\sgn(w_i^0), \forall i\in \cJ^0$, then \eqref{kkt} holds
   for $\hbmu_n$ in \eqref{hmuj}.
   Thus,  $\hbmu_n$ is a LAD-FSA solution.
   Define
  $$\cR_{\lmtwo}\equiv \{\widehat\cJ=\cJ^0 \}\cap \{\sgn(\widehat w_i^F)=\sgn(w_i^0), ~\forall i\in \cJ^0\}$$
   Then $\cR_{\lmtwo}$ holds if
    \begin{subequations}
     \begin{empheq}[left=\empheqlbrace]{align}
      &\sgn(\widehat w_i)=\sgn(w_i^0) &        \forall i\in \cJ^0\label{ktt-1}\\
      &   \arrowvert\sum_{k\in \widehat\cB_{j(i)}} \nolimits\sgn(y_k-\hmu_i)\arrowvert<2\lmtwo   & {\rm if}~i\notin \cJ^0 \label{kkt-2}
     \end{empheq}
    \end{subequations}
  It is easy to verify that 
  $\sgn(\widehat w_i)=\sgn(w_i^0), \forall i\in \cJ^0$ holds
if
\bel{hmuj3}
|\sgn(\widehat w_i)(w_i^0-\widehat w_i)|<|w_i^0|,~{\rm for~}i\in \cJ^0
\eel
Plug  $\hbmu_n$ \eqref{hmuj}  into  \eqref{hmuj3} and \eqref{kkt-2}, and then use the triangle inequality,
we know that
$\cR_{\lmtwo}$ holds if
\begin{equation}\label{Rlam2n-1}
 \begin{array}{l}
  \max_{i\in \cJ^0}|(b_{j(i)}^0)^{-1}\sum_{k\in \cB^0_{j(i)}} \sgn(\veps_k)-
  (b_{j(i-1)}^0)^{-1}\sum_{k\in \cB^0_{j(i-1)}} \sgn(\veps_k)|/w_i^0+ \\
        \quad\max_{i\in \cJ^0} |(b_{j(i)}^0)^{-1}\widehat h_{i}-(b_{j(i-1)}^0)^{-1}\widehat h_{i-1}|/w_i^0+
    \max_{i\in \cJ^0}|c_{j(i)}^0-c_{j(i-1)}^0|/w_i^0 \\ <2f(0).
  \end{array}
 \end{equation}
 and
 \begin{equation}\label{Rlam2n-2}
 \max_{i\notin \cJ^0} |\sgn(\veps_i)-\sgn(\veps_{i-1})+\widehat h_{i}-\widehat h_{i-1}|<4\lmtwo.
  \end{equation}
  We have $$\bE[\sgn(\veps_i)-\sgn(\veps_{i-1})]=0$$
  and $$\Var[\sgn(\veps_i)-\sgn(\veps_{i-1})]=2 ~{\rm for ~}2\le i\le n$$
  and for $2\le i_l, i_2 \le n$,
   $$\Cov(\sgn(\veps_{i_1})-\sgn(\veps_{{i_1}-1}), \sgn(\veps_{i_2})-\sgn(\veps_{{i_2}-1}) )=-1, 0~{\rm for~} |{i_1}-{i_2}|=1, ~{\rm otherwise}.$$
Suppose $d_i^*$ are independent copies of $N(0,2)$.
Then we have
  $$\begin{array}{ll}
  \bP(I_4)&\equiv \bP(\max_{i\notin \cJ^0} |\sgn(\veps_i)-\sgn(\veps_{i-1})|>2\lmtwo)\\
   &\quad\le \bP(\max_{i\notin \cJ^0}|d_i^*|> 2\lmtwo)\\
   &\quad\le 2\exp\{-4\lmtwo^2+\log|\cJ_0^c|\}
   \end{array}
  $$
  where we get the first ``$ \le $''  using Slepian's inequality, the second  ``$\le$''
  using Chernoff's bound.
  Then $\bP(I_4)=o(1)$  if conditions in (B1) hold.
   Define
   $$X_i=(2f(0)b_{j(i)}^0)^{-1}\sum_{k\in \cB^0_{j(i)}} \sgn(\veps_k)-(2f(0)b_{j(i-1)}^0)^{-1}\sum_{k\in \cB^0_{j(i-1)}} \sgn(\veps_k)| ~\forall i\in \cJ^0.$$
    Then $\bE[X_i]=0$ and
    $\max_{i\in \cJ^0}\Var[X_i]\le (2f(0)b_{j(i)}^0)^{-1}$.
    Consider independent copies $X_i^*\sim N(0, (2f(0)b_{j(i)}^0)^{-1} ), i\in \cJ^0$. We have
    $$
    \begin{array}{ll}
 \bP(I_1)&\equiv \bP(\max_{i\in \cJ^0}|(b_{j(i)}^0)^{-1}\sum_{k\in \cB^0_{j(i)}} \sgn(\veps_k)-
  (b_{j(i-1)}^0)^{-1}\sum_{k\in \cB^0_{j(i-1)}} \sgn(\veps_k)|>2f(0)a_n/3)\\
  &\quad \le \bP(\max_{i\in \cJ^0}|X_i^*|>a_n/3)\le 2\exp\{-2b^0_{\min}f^2(0)a_n^2/9+\log|\cJ^0|\}.
  \end{array}
$$
Thus $\bP(I_1)=o(1)$ if conditions in (B2) holds.
Since  $\max_{i\in \cJ^0}|c_{j(i)}^0-c_{j(i-1)}^0| \le 2\lmtwo /b_{\min}^0$,
from (B3),
$$
\bP(I_2)\equiv \bP(\max_{i\in \cJ^0}|c_{j(i)}^0-c_{j(i-1)}^0|> 2f(0) a_n/3)=0.
$$
 Furthermore, we have $$
\bP(I_5)\equiv \bP( \max_{i\notin \cJ^0} |\widehat h_i-\widehat h_{i-1}|>2\lmtwo)=0.
 $$
From \eqref{equicontinuity}, we have
$$
    \begin{array}{ll}
\bP(I_3)&\equiv \bP(\max_{i\in \cJ^0} |(w_i^0b_{j(i)}^0)^{-1}\widehat h_{i}-(w_i^0b_{j(i-1)}^0)^{-1}\widehat h_{i-1}|> 2f(0)/3)\\
&\le \bP(\max_{i\in \cJ^0} |(b_{j(i)}^0)^{-1/2}\widehat h_{i}-(b_{j(i-1)}^0)^{-1/2}\widehat h_{i-1}|> 2f(0)(b_{\min}^0)^{1/2}a_n/3)\\
&=o(1).
 \end{array}
$$
Then from \eqref{Rlam2n-1} and \eqref{Rlam2n-2}, we get
$$
\bP(\cR_{\lmtwo}^c)\le \bP(I_1)+\bP(I_2)+\bP(I_3)+\bP(I_4)+\bP(I_5) \to 0 ~{\rm when~} n\to \infty.
$$
 $\Box$

\vspace{15pt}

 \noindent{\bf Proof of Theorem \ref{theorem-consistency-flsa}}

  \noindent From Theorem \ref{theorem-consistency-fsa}, we know that if (A1) and (B1-B3) hold,
  the  LAD-FLSA
  can choose all jumps with probability  $1$.
   Thus, we can prove the main results based on the true block partition.
 By the KKT, $\hbnu_n$ is a LAD-FLSA solution if and only if
 \bel{kkt-flsa}
 \left\{
  \begin{array}{ll}
  \sum_{k\in \cB^0_{j}} \sgn(y_k- \hnu_j)+ \widehat b_j \widehat c_j= \lmone \widehat b_j \sgn(\hnu_j) &{\rm if}~\hnu_j\neq 0 \\
  |\sum_{k\in \cB^0_{j}} \sgn(y_k-\hnu_j)|+ \widehat b_j \widehat c_j|<\lmone \widehat b_j   &  {\rm if}~\hnu_j= 0
  \end{array}
  \right..
 \eel
Let  $\hnu_j$ and $\nu_j^0$ satisfy
      \begin{equation} \label{hnuj}
     \left\{
     \begin{array}{ll}
     \hnu_j=\nu_j^0+(2f(0)b_{j}^0)^{-1}\left(\sum_{i\in \cB_{j}^0} \sgn(\veps_i)+ b^0_j c_j^0-\lmone b_j^0 \sgn(\nu_j^0)+\widehat h_{j}\right) &        \quad \forall~ j\in \cK^0 \\
       \hnu_j=0   & \quad\forall~ j\in \cK^0
     \end{array}
     \right..
    \end{equation}
Here, by abuse of notation,
$\widehat h_{j}$ is the remainder term with the stochastically equicontinunity,
\bel{equicontinuity2}
|(b_j^0)^{-1/2}\widehat h_{j}|=O_p(1), ~\forall {1\le i\le  n}.
\eel
In fact, $\widehat h_{j}=2f(0)b^0_{j} (\hnu_j-\nu_j^0)+\sum_{ i\in \cB^0_{j}}\bE[r_i^j] +\sum_{i\in \cB^0_{j}}\varsigma_i^j$,
with $r_i^j=\sgn(\veps_i-(\hnu_j-\nu_j^0) )-\sgn(\veps_i)$ and $\varsigma_i^j=r_i^j-\bE_{\bveps}[r_i^j]$ for $i\in \cB^0_j$
and $1\le j\le J_0+1$.
If $\{\sgn(\hnu_j)=\sgn(\nu_j^0), \forall ~j\in \cK^0\}$, then $\bnu_n$ in \eqref{hnuj}
satisfies {kkt-flsa}, and therefore is  a LAD-FLSA solution.
 Define an event
$$\cR_{n}=\cR(\lmone,\lmtwo)\equiv\{\widehat{\cK}=\cK^0\}\cap\{\sgn(\hnu_j)=\sgn(\nu_j^0), \forall ~j\in \cK^0\}.$$
  Then $\cR_n$ holds if
 \bel{flsa-kkt2}
\left\{
  \begin{array}{ll}
     \{\sgn(\hnu_j)=\sgn(\nu_j^0),  & \forall ~j\in \cK^0 \\
 |\sum_{k\in \cB^0_{j}} \sgn(y_k-\hnu_j)|+ \widehat b_j \widehat c_j|<\lmone \widehat b_j & \forall ~j\notin \cK^0  \\
  \end{array}
  \right..
\eel
We can verify that $\sgn(\widehat \nu_j)=\sgn(\nu_j^0), \forall j\in \cK^0$ holds
if
$|\sgn(\hnu_j)(\nu_j^0-\widehat \nu_j)|<|\nu_j^0|,~{\rm for~}j\in \cK^0.
$
Therefore, from \eqref{hnuj} and \eqref{flsa-kkt2},   $\cR_n$ holds if
 \bel{hnuj3}
 \left\{
   \begin{array}{ll}
   |\sum_{i\in \cB^0} \sgn(\veps_i) +b_j^0c_j^0-\lmone b_j^0 \sgn(\nu_j^0)+\widehat h_j|<2f(0)b_j^0|\nu_j^0|& \forall~ j\in \cK^0\\
   |\sum_{i\in \cB^0} \sgn(\veps_i) +b_j^0c_j^0+\widehat h_j|<\lmone b_j^0& \forall ~j\notin \cK^0
   \end{array}
 \right..
 \eel
Thus we have
\bel{R1n}
 \begin{array}{ll}
 \bP(\cR^c)
 &\le \bP(\max_{j\in \cK^0} |\sum_{i\in \cB_j^0} \sgn(\veps_i)|> 2f(0) \min_{j\in\cK^0} b_j^0 \min_{j\in\cK^0} |\nu_j^0|/4)\\
   &+ \bP(\max_{j\in \cK^0} |c_j^0| > 2f(0) \min_{j\in\cK^0}|\nu_j^0|/4 )\\
   &+\bP(\max_{j\in \cK^0} |\lmone  \sgn(\nu_j^0)|> 2f(0) \min_{j\in\cK^0} |\nu_j^0|/4 )\\
   &+ \bP(\max_{j\in \cK^0}|\widehat h_j/(b_j^0\nu_j^0)|>f(0) /2)\\
   &+\bP(\max_{j\notin \cK^0} |\sum_{i\in \cB_j^0} \sgn(\veps_i)|>\lmone \min_{j\in\cK^0} b_j^0/3 )\\
    &+\bP(\max_{j\notin \cK^0} |c_j^0|> \lmone/3)\\
    &+\bP(\max_{j\notin \cK^0}|\widehat h_j/b_j^0|>\lmone /3)\\
   &\equiv\bP(S_1)+\bP(S_2)+\bP(S_3)+\bP(S_4)+\bP(S_5)+\bP(S_6)+\bP(S_7).
\end{array}
 \eel
Let $Z_j=\sum_{i \in \cB_j^0} \sgn(\veps_i)/b_j^0$. Then $\bE[Z_j]=0$ and $\Var(Z_j)=1/b_j^0$.
Then $Z_j$s are independent sub-Gaussian.
From (C3), we have
$$ P(S_1) \le 2K_0 \exp\{-b_{\min}^0f^2(0)\rho_n^2/8\}=o(1).$$
We can verify $P(S_2)=o(1)$ from (C4), $P(S_3)=o(1)$ from (C5) and $P(S_6)=o(1)$ from (C2)
From (C1), $$P(S_5)\le 2(J_0-K_0)\exp\{-b_{\min}^0\lmone^2/32\}=o(1).$$
 Furthermore, we have $P(S_7)=o(1)$ and  $P(S_4)= o(1)$.
 From \eqref{R1n}, we have $P(\cR) \to 1$ when $n\to \infty$, which completes the proof.
$\Box$

 \vspace{15pt}
 The rest of the Appendix are presented to prove  Theorem \ref{thm-unbiased-fl}.

Recall that $\bw$ is the jump coefficients vector with $w_i=\mu_i-\mu_{i-1}$ for $2\le i\le n$
and $\bnu=(\nu_1,\cdots, \nu_J)'$ is the block coefficients factor.
 From Proposition 3 in Rosset and Zhu (2007), we know that the following results of the LAD-FLSA solution.
 \begin{lemma}\label{piecewise}
 \begin{itemize}
\item[(i)] For any $\lm_1=0$, there exists a set of values of $\lm_2$,
$$0=\lm_{2,0}<\lm_{2,1}<\cdots<\lm_{2,m_2}<\lm_{2,m_2+1}=\infty$$ such
that $\widehat \bw(0, \lm_{2,k})$ for $1\le k\le m_2$ is not uniquely defined, the set of optimal solutions of
each $\lm_{2,k}$  is a straight line in $\cR^n$, and for any $\lm_2\in (\lm_{2,k}, \lm_{2,k+1})$, the
solution  $\widehat \bw(0, \lm_2)$ is constant.
\item[(ii)] For above $\lm_{2,k}, 1\le k\le m_2$,  there exists a set of values of $\lm_1$,
$$0=\lm_{1,0}<\lm_{1,1}<\cdots<\lm_{1,m_2}<\lm_{1,m_2+1}=\infty$$ such
that $\hbnu(\lm_{1,j}, \lm_{2,k})$ for $1\le j\le m_1$ is not uniquely defined,
the set of optimal solutions of
each $\lm_{1,j}$ is a straight line in $\cR^n$, and for any $\lm_1\in (\lm_{1,j}, \lm_{1,j+1})$, the
solution  $\hbnu(\lm_{1}, \lm_{2,k})$ is constant.
\end{itemize}
\end{lemma}
In Lemma~\ref{piecewise}, if we define  $\lm_{2,k}, 1\le k \le m_2$ from  (i)  as the transition points for $\bw$ and
 $\cN_{0, \lm_2}$ as the set of
 $\by\in \cR^n$ such that $\lm_{2}$ is a transition point for $\bw$,
 then the jumps set  $\cJ(0,\lm_2)$ only changes at those $\lm_{2,k}$'s.
Furthermore, Let $\lm_2=\lm_{2,k}$ for some $1\le k \le m_2$.
 If we can also define $\lm_{1,j}, 1\le j \le m_1$ from (ii) as the transition points for $\bnu$ and
 $\cN_{\lm_{1}, \lm_{2}}$ as the set of
 $\by\in \cR^n$ such that $\lm_{1}$ is a transition point for $\bnu$,
then the set of nonzero blocks,  $\cK(\lm_{1},\lm_{2})$,  only changes at  $\lm_{1,j}$'s or $\lm_{2,k}$'s,
and $\cN_{\lm_1,\lm_2}$, is in a finite collection of hyperplanes in $\cR^n$.
From Lemma \ref{piecewise}, we know that for any given
$\by\in\cR^n/\cN_{\lm_{1}, \lm_2}$, $\hbmu(\lm_{1},\lm_2)$ is fixed and then $\{1, 2,
\cdots, n\}$ is divided into two sets, $\cE_{\by,\lm_{1},\lm_2}$ and
$\cE_{\by,\lm_{1},\lm_2}^c$, where $\cE_{\by,\lm_{1},\lm_2}=\{1\le i\le n:~ y_i-\hnu_{j(i)}=0, \hnu_{j(i)}\neq 0\}
$
and $j(i)$ specifies the block where $\hmu_i$ stays.
Thus, we have
\bel{event}
 |\cE_{\by,\lm_1,\lm_2}|=|\cK(\lm_1,\lm_2)|.
\eel

\begin{lemma}\label{continuous}
For any $\lm_1>0$ and $\lm_2>0$,  if $\by\in\cR^n/\cN_{\lm_{1}, \lm_2}$,
 $\hnu(\lm_1,\lm_2,\by)$ is a
continuous function of $\by$, and then $\cE_{\by,\lm_1,\lm_2}$ is locally
constant.
\end{lemma}

\noindent {\bf Proof of Lemma~\ref{continuous}}

\noindent
Let $L(\mu, \by)$ denote the function
$$L(\bmu, \by)=\sum_{i=1}^n|y_i-\mu_i|+\lm_1\sum_{i=1}^n|\mu_i|+\lm_2\sum_{i=2}^n|\mu_i-\mu_{i-1}|.
$$
 Since
$\by\in\cR^n/\cN_{\lm_1,\lm_2}$,  $\widehat \bnu$
does not change from Lemma \ref{piecewise}.
 For any $\by_0\in\cR^n/\cN_{\lm_1,\lm_2}$, and for any sequence $\{\by_m\}$ such that
 $\{\by_m\}\to \by_0$, we want to prove that
 $\hbnu(\lm_1,\lm_2, \by_m)\to \hbnu(\lm_1,\lm_2, \by_0)$. It is equivalent to
 prove  $\hbmu(\lm_1,\lm_2, \by_m)\to \hbmu(\lm_1,\lm_2, \by_0)$.
 Because $\|\hbmu(\lm_1,\lm_2,\by)\|_1\leq\|\hbmu(0,0,\by)\|_1=\|\by\|_1$,
  $\hbmu(\lm_1,\lm_2,\by)$    is bounded. Thus we only need to check that for every
   converging subsequence of $\{\by_{m}\}$, say
  $\{\by_{m_k}\}$, we have $\hbmu(\lm_1,\lm_2, \by_{m_k})\to \hbmu(\lm_1,\lm_2, \by_0).$ Suppose that
$\hbmu(\lm_1,\lm_2, \by_{m_k})\to \check\bmu(\lm_1,\lm_2)$ when $m_k\to \infty$.
Let $\Delta(\bmu,\by,\by')=L(\bmu, \by)-L(\bmu, \by')$.
On the one hand, we have
\bes
&&L(\hbmu(\lm_1,\lm_2, \by_{0}),\by_{0})\\
&=& L(\hbmu(\lm_1,\lm_2, \by_0),\by_{m_k})+\Delta(\hbmu(\lm_1,\lm_2, \by_0),\by_{0},\by_{m_k}) \\
&\ge& L(\hbmu(\lm_1,\lm_2, \by_{m_k}),\by_{m_k})+\Delta(\hbmu(\lm_1,\lm_2, \by_0),\by_{0},\by_{m_k})\\
&=&  L(\hbmu(\lm_1,\lm_2, \by_{m_k}),\by_{0})+\Delta(\hbmu(\lm_1,\lm_2, \by_{m_k}),\by_{m_k},\by_{0})
    +\Delta(\hbmu(\lm_1,\lm_2, \by_0),\by_{0},\by_{m_k}).
\ees
On the other hand, we have
\bes &&\Delta(\hbmu(\lm,
\by_{m_k}),\by_{m_k},\by_{0})
    +\Delta(\hbmu(\lm_1,\lm_2, \by_0),\by_{0},\by_{m_k})\\
&=&\sum_{i=1}^n[|y_{i,m_k}-\hmu_i(\lm_1,\lm_2, \by_{m_k})|-
         |y_{i,0}-\hmu_i(\lm_1,\lm_2, \by_{m_k})|\\
         &&+|y_{i,0}
          -\hmu_i(\lm_1,\lm_2, \by_{0})|-
         |y_{i,m_k}-\hmu_i(\lm_1,\lm_2, \by_{0})|]\\
&\le&  2\sum_{i=1}^n|y_{i,m_k}-y_{i,0}|\to 0 ~{\rm when~} k\to \infty
\ees
Thus $
   L(\hbmu(\lm_1,\lm_2, \by_{0}),\by_{0})
    \ge \lim_{k\to \infty} L(\hbmu(\lm_1,\lm_2, \by_{m_k}),\by_{0})
      = L(\check\bmu(\lm_1,\lm_2,\by_{0}),\by_{0})$. Since $\hbmu(\lm_1,\lm_2, \by_{0})$ is
the unique minimizer of $L(\bmu,\by_{0})$, we have
$\check\bmu(\lm_1,\lm_2,\by_{0})=\hbmu(\lm_1,\lm_2, \by_{0})$.
 $\Box$

\vspace{15pt}

 \noindent{\bf Proof of Theorem~\ref{thm-unbiased-fl}}

  \noindent
From \eqref{event} and Lemma \ref{continuous},
 there exists $\eps>0$ such that $\by\in {\rm Ball}(\by, \eps)$,  $\cE_{\by,\lm_1,\lm_2}$ stays the same when
neither $\lm_1$ and $\lm_2$ is a transitional point.
Thus, $\partial \hnu_{j(i)}(\lm_1,\lm_2)/\partial y_i =1$ if $i\in \cE_{\by,\lm_1,\lm_2}$ and  $\partial \hnu_{j(i)}(\lm_1,\lm_2)/\partial y_i =0$ if $i\notin \cE_{\by,\lm_1,\lm_2}$.
  Overall, we have $\sum_{i=1}^n \partial\hnu_{j(i)}(\lm_1,\lm_2)/\partial y_i=|\cE_{\by,\lm_1,\lm_2}|= |\widehat \cK(\lm_1,\lm_2)|$ for
 $\by \in \cN_{\lm_1,\lm_2}$.
 Since $\cN_{\lm_1,\lm_2}$ is in a collection of finite
 hyperplanes, we can obtain the conclusion by taking the expectation. $\Box$

\renewcommand{\baselinestretch}{1.0}

\newpage
\begin{figure}[htp]
\centering
 $$\scalebox{0.3}[0.5]{\includegraphics{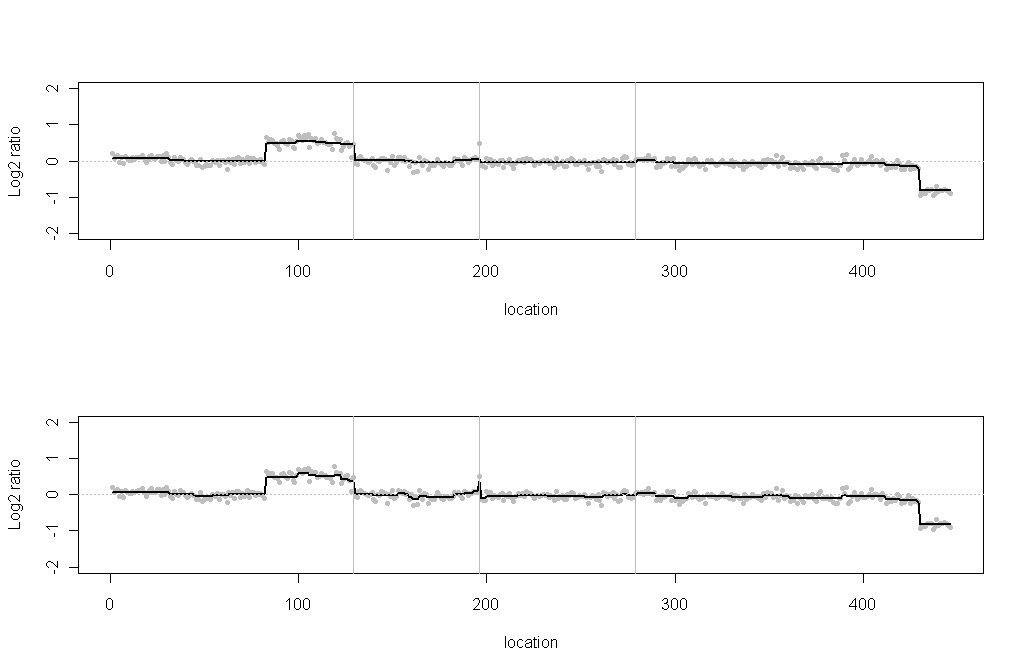}}$$
\caption{Copy Number data set from the GM 13330  BAC CGH array. The top and bottom panels give
outputs from the LAD-FLSA and LS-FLSA, respectively.
Both observed data (gray dots) and estimates (dark solid lines) from chromosome 1--4 are plotted .
Data from different chromosomes are separated by gray vertical lines. }\label{fig:signal example}
\end{figure}

\begin{figure}[htp]
\centering
 $$\scalebox{0.7}[0.7]{\includegraphics{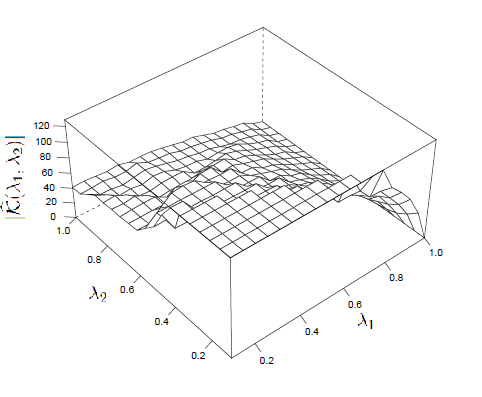}}$$
\caption{The estimated degrees of freedom of LAD-FLSA
for every combined $\lm_1$ and $\lm_2$ for the
 chromosome 1 data from the GM 13330 BAC array.
}\label{fig:signal example}
\end{figure}

\begin{figure}[htp]
\centering
$$\scalebox{0.4}[0.4]{\includegraphics{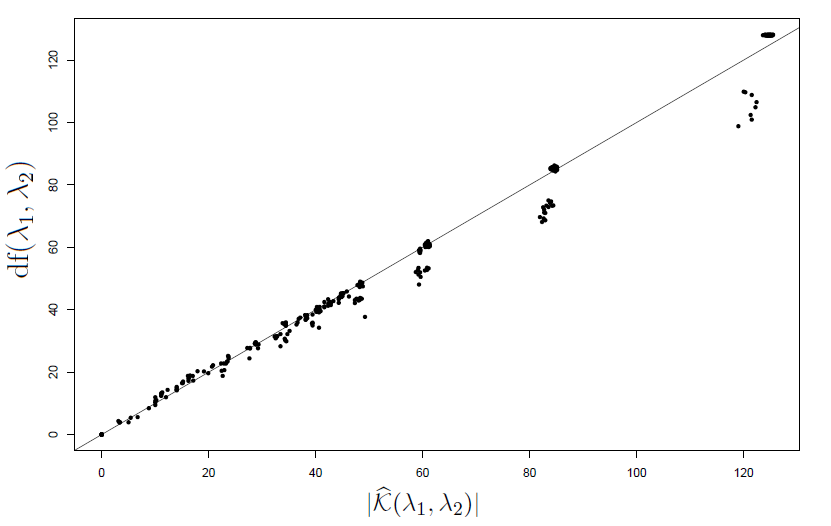}}$$
\caption{Hypothetical model from 500 Monto Carlo simulations of 129
markers for chromosome 1  data from the GM 13330 BAC array. It shows
that the estimated number of nonzero blocks $|\cK(\lm_1,\lm_2)|$
 is very close to the true $D(\mathcal{M}_{\lm_1,
\lm_2})$ using the 45 degree line.}\label{fig:CGHdf}
\end{figure}

\begin{figure}[htp]
\centering
$$\scalebox{0.6}[0.6]{\includegraphics{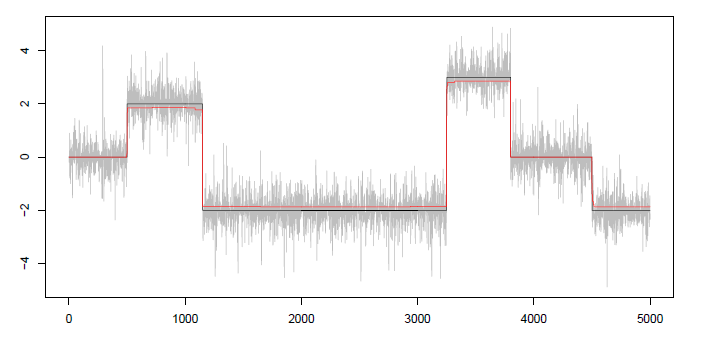}}$$
\caption{Example of observed data (grey) with true hidden signals
 (black) and LAD-FLSA estimates (red). There are $6$ blocks
 with $4$ nonzero ones. Random noise $\veps_i$'s are generated
  from double exponential distributions with center $0$ and
  scale $0.5/\sqrt{2}$. }\label{fig:signaldata}
\end{figure}

\begin{table} [h] 
 \begin{center}
   \caption {Simulation results for Section \ref{sec-sim-con}.
   }
  {\small  \begin{tabular} {c c c  c c c    c c c  }\label{sim results} \\ \hline\hline
                         &  &  &
                         \multicolumn{3}{c}  {$n=1000 $}
                         & \multicolumn{3}{c}  {$n=5000$}
              \\
                         \cline{2-5} \cline{6-9}
     $\epsilon_i$ & $\sigma$ & {\bf Model}
                            & LARE$^1$& CFR+6$^2$& JUMP$^3$
                            & LARE    &CFR+6     &  JUMP   
                             \\ \hline\hline
                             \multirow{6}{*}{\rotatebox{90}{\mbox{Normal}}}
&    & LAD-FLSA   &0.197  &89\%(17\%)  &7.12(1.32)         &0.173  & 82\%(15\%) &7.24(1.33)   \\
& \raisebox{1.3ex}[0pt]{1.0}
    & LS-FLSA   & 0.035  &18\%(3\%)  & 7.82(1.47)       &0.021  &5\%(0\%)  &7.7(1.48)\\
 \cline{3-9}

&    & LAD-FLSA   &0.098  &97\%(32\%)  & 5.59 (0.75)      &0.087  & 96\%(22\%) &5.54(0.72) \\
& \raisebox{1.3ex}[0pt]{0.5}
      & LS-FLSA   & 0.016  &48\%(13\%) & 5.68(0.74)     &0.007  &57\%(7\%)  &5.61(0.74)\\
 \cline{3-9}

&    & LAD-FLSA  & 0.019  &100\%(93\%)  &5.00(0.00)       &0.017  & 100\%(94\%)  &5.00(0.00) \\
& \raisebox{1.3ex}[0pt]{0.1}
      & LS-FLSA   & 0.013  &100\%(93\%)  & 5.00(0.00)      &0.003  &100\%(94\%)  &5.00(0.00) \\

 \cline{3-9}
        \hline
 \multirow{6}{*}{\rotatebox{90}{\mbox{Double Exp.}}}
&    & LAD-FLSA  &0.154  &88\% (22\%)  &7.42(1.54)      &0.128 & 89\%(25\%) &7.18(1.35) \\
& \raisebox{1.3ex}[0pt]{1.0}
     & LS-FLSA   & 0.031  &12\%(0\%)  & 7.42(1.42)     &0.021  &3\%(1\%)  &7.31(1.43)\\
 \cline{3-9}

&    & LAD-FLSA  &0.077  &97\%(34\%)  &5.95(0.90)     &0.064 & 100\%(41\%) &5.74(0.86) \\
& \raisebox{1.3ex}[0pt]{0.5}
    & LS-FLSA   & 0.016  &57\%(12\%)  &5.73(0.78)     &0.007  &62\%(19\%)  &5.68(0.87)\\
 \cline{3-9}

&    & LAD-FLSA  & 0.015  &100\%(97\%)  &5.00(0.00)       &0.013  & 100\%(90\%)  &5.00(0.00) \\
& \raisebox{1.3ex}[0pt]{0.1}
     & LS-FLSA   & 0.013  &100\%(97\%)  & 5.00(0.00)      &0.003  &100\%(89\%)  &5.00(0.00) \\

 \cline{3-9}
        \hline
 \multirow{6}{*}{\rotatebox{90}{\mbox{Cauchy}}}
&    & LAD-FLSA  &0.048  &87\%(56\%)  &6.12(1.07)      &0.029  & 82\%(45\%) &6.14(0.95) \\

& \raisebox{1.3ex}[0pt]{1.0}
      & LS-FLSA   & 0.239  &17\%(4\%)  & 16.37(5.38)     &0.275  &2\%(0\%)  &59.41(14.38)\\
 \cline{3-9}

&    & LAD-FLSA  &0.028  &99\%(70\%)  &5.56(0.86)      &0.015  & 87\%(66\%) &5.59(0.78) \\

& \raisebox{1.3ex}[0pt]{0.5}
    & LS-FLSA   & 0.120 &39\%(17\%)  & 10.67(3.62)    &0.132  &15\%(3\%)  &32.05(8.51)\\
 \cline{3-9}

&    & LAD-FLSA  & 0.007  &95\%(92\%)  &5.18(0.46)       &0.003  & 96\%(87\%)  &5.17(0.40) \\
& \raisebox{1.3ex}[0pt]{0.1}
     & LS-FLSA   & 0.029  &94\%(78\%)  & 6.30(1.34)      &0.023  &87\%(49\%)  &10.14(3.13) \\
 \cline{3-9}

        \hline\hline
  \multicolumn{9}{l} {\footnotesize{NOTE 1: LARE is the least absolute relative ratio defined in \eqref{eq:lare}.}}\\
  \multicolumn{9}{l} {\footnotesize{NOTE 2: CFR+6 is the ratio of recovering $\hmu^0$
                       correctly or plus at most six additional false posives}}\\
  \multicolumn{9}{l} {\footnotesize{   \quad\quad\quad\quad                   (correctly fitted ratio). }}\\
  \multicolumn{9}{l} {\footnotesize{NOTE 3: JUMP is the average number (standard deviation)  of the number of jumps.}}
   \end{tabular}}
  \end{center}
  \end{table}

\end{document}